\documentclass[twocolumn,superscriptaddress,floatfix,preprintnumbers, nofootinbib,hyperref]{revtex4-2} 
\pdfoutput=1
\usepackage[colorlinks=true,breaklinks=true]{hyperref}
\usepackage[normalem]{ulem}
\usepackage[utf8]{inputenc}
\hypersetup{allcolors=[rgb]{0.0 0.0 0.6},linkcolor=[rgb]{0.75 0.05 0.05}}
\usepackage{amsmath,amssymb}
\usepackage{epsfig}  
\usepackage{graphicx}   
\usepackage{slashed}       
\usepackage{tikz}
\usepackage{color}
\usepackage{url}
\usepackage{color}
\usepackage{multirow}
\usepackage{comment}
\usepackage{amssymb}
\usepackage{bm}

\DeclareMathOperator{\cm}{cm}
\DeclareMathOperator{\GeV}{GeV}
\DeclareMathOperator{\neV}{neV}
\DeclareMathOperator{\eV}{eV}

\DeclareMathOperator{\MeV}{MeV}

\DeclareMathOperator{\s}{s}

\DeclareMathOperator{\kpc}{kpc}

\newcommand{\bE}{{\bf E}}
\newcommand{\bB}{{\bf B}}

\newcommand{\beq}{\begin{equation}}
\newcommand{\eeq}{\end{equation}}

\newcommand{\fermi}{\textit{Fermi}}

\definecolor{forestgreen}{rgb}{0.13, 0.545, 0.13}

\allowdisplaybreaks

\bibpunct{[}{]}{,}{n}{}{,}

\begin{document}

\preprint{LAPTH-031/23}

\title{Uncovering axion-like particles in supernova gamma-ray spectra}

\author{Francesca Calore}
\email{calore@lapth.cnrs.fr}
\affiliation{LAPTh, CNRS, F-74000 Annecy, France}

\author{Pierluca Carenza}
\email{pierluca.carenza@fysik.su.se}
\affiliation{The~Oskar~Klein Centre,~Department~of~Physics,~Stockholm~University,~Stockholm~106 91,~Sweden}

\author{Christopher Eckner}
 \email{eckner@lapth.cnrs.fr}
\affiliation{LAPTh, CNRS, F-74000 Annecy, France}
\affiliation{LAPP, CNRS, F-74000 Annecy, France}

\author{Maurizio Giannotti}
\email{mgiannotti@barry.edu}
\affiliation{Physical Sciences, Barry University, 11300 NE 2nd Ave., Miami Shores, FL 33161, USA}

\author{Giuseppe Lucente}
\email{giuseppe.lucente@ba.infn.it }
\affiliation{Dipartimento Interateneo di Fisica ``Michelangelo Merlin'', Via Amendola 173, 70126 Bari, Italy.}
\affiliation{Istituto Nazionale di Fisica Nucleare - Sezione di Bari, Via Orabona 4, 70126 Bari, Italy.}

\author{Alessandro Mirizzi}
\email{alessandro.mirizzi@ba.infn.it }
\affiliation{Dipartimento Interateneo di Fisica ``Michelangelo Merlin'', Via Amendola 173, 70126 Bari, Italy.}
\affiliation{Istituto Nazionale di Fisica Nucleare - Sezione di Bari, Via Orabona 4, 70126 Bari, Italy.}

\author{Francesco Sivo}
\email{francesco.sivo@ba.infn.it }
\affiliation{Dipartimento Interateneo di Fisica ``Michelangelo Merlin'', Via Amendola 173, 70126 Bari, Italy.}
\affiliation{Istituto Nazionale di Fisica Nucleare - Sezione di Bari, Via Orabona 4, 70126 Bari, Italy.}

\smallskip

\begin{abstract}
A future Galactic Supernova (SN) explosion can lead to a gamma-ray signal induced by ultralight Axion-Like Particles (ALPs) thermally produced in the SN core and converted into high-energy photons in the Galactic magnetic field. 
The detection of such a signal is in the reach of the Large Area Telescope aboard the \emph{Fermi} Gamma-Ray Space Telescope. The observation of gamma-ray emission from a future SN has a sensitivity to $g_{a\gamma}\gtrsim 4\times 10^{-13}~\GeV^{-1}$ for a SN at fiducial distance of $10~\kpc$ and would allow us to reconstruct the ALP-photon coupling  within a factor of $\sim2$, mainly due to the uncertainties on the modeling of the Galactic magnetic field.
\end{abstract}

\maketitle

\section{Introduction}

The detection of gamma-ray signals in coincidence with a  Galactic core-collapse (CC) Supernova (SN) explosion has been pointed out as smoking-gun signature associated with the emission of novel particles, like axions~\cite{Grifols:1996id,Brockway:1996yr,Payez:2014xsa,Jaeckel:2017tud,Calore:2020tjw,Caputo:2021rux,Hoof:2022xbe,Muller:2023vjm}, heavy sterile 
neutrinos~\cite{Oberauer:1993yr,Fuller:2008erj,Arguelles:2016uwb}, or dark photons~\cite{DeRocco:2019njg}. 
In this context, one of the most studied possibilities is the case  of ultralight Axion-Like Particles (ALPs), coupled with photons through the coupling $g_{a\gamma}$, that  would be  thermally produced in the SN core via Primakoff process~\cite{Grifols:1996id,Brockway:1996yr}.   These ALPs are expected to freely escape from  the SN core and travel through the Galactic magnetic field, converting into photons, and potentially generating an unexpected gamma-ray burst simultaneously with the neutrino signal. 

A notable application of this mechanism is the strong bound, ${g_{a\gamma} \lesssim 4.2 \times 10^{-12}~\GeV^{-1}}$ for ${m_a < 4 \times 10^{-10}}$~eV~\cite{Hoof:2022xbe},  from the lack of a gamma-ray signal in the Gamma-Ray Spectrometer on the Solar Maximum Mission in coincidence with the neutrino signal from SN 1987A  (see also~\cite{Payez:2014xsa} for a previous estimation of the bound).
The physics potential of current gamma-ray detectors has been extensively explored in the past few years.
In particular, it was shown that if a Galactic SN were to explode during the lifetime of the Large Area Telescope aboard the {\it Fermi}~satellite (hereinafter {\it Fermi}-LAT), one could 
probe the ALP parameter space  significantly below the previous constraints, with a sensitivity down to ${g_{a\gamma}\gtrsim 2\times 10^{-12}~\GeV^{-1}}$ for ultralight ALPs and a SN placed in the Galactic center~\cite{Meyer:2016wrm}.
Furthermore, a search for gamma-ray bursts from extragalactic SNe with {\it Fermi}-LAT  has yielded the limit ${g_{a\gamma} < 2.6 \times 10^{-11}~\GeV^{-1}}$, for ALP masses
${m_a < 3 \times 10^{-10}~\eV}$, under the assumption of at least one SN occurring in the detector field of view~\cite{Meyer:2020vzy}.
Finally, the cumulative emission of ALPs from all past CC SNe in the universe
would produce a diffuse ALP flux, whose conversion into photons in the Galactic magnetic field can lead to a potentially detectable diffuse gamma-ray flux
at MeV energies. 
Using recent measurements of the diffuse gamma-ray flux observed by {\it Fermi}-LAT
one can set the bound  $g_{a\gamma} \lesssim 3.8 \times 10^{-11}$~GeV$^{-1}$ for $m_a \ll 10^{-11}$~eV~\cite{Calore:2020tjw,Calore:2021hhn}.

In this paper, we explore the sensitivity of \fermi-LAT
to the detection of a gamma-ray signal induced by ALPs emitted from a Galactic SN. 
In particular, we will demonstrate that a detection of such a signal, in coincidence with the SN neutrino burst, would represent a hint for new physics, providing a unique opportunity to reveal the existence of ALPs and precisely reconstruct their properties, such as the ALP-photon coupling and the average energy of their spectrum.\footnote{Following a similar rationale,  some of the authors have analyzed the sensitivity to heavy, MeV-scale, ALPs from SNe in  Ref.~\cite{Muller:2023vjm}.} 

Our work follows this structure. In Sec.~\ref{sec:alp_gamma_conv}, we characterize the initial SN ALP flux, the ALP-photon conversion in the Galactic magnetic field and the observable gamma-ray flux. 
In Sec.~\ref{sec:fermi_analysis}, we assess the capability of \emph{Fermi}-LAT to reconstruct the ALP parameters after the observation of a gamma-ray signal from a Galactic SN explosion. The results of this analysis are discussed in Sec.~\ref{sec:results}. Finally, in Sec.~\ref{sec:conclusions} we summarize and conclude.

\section{ALP-induced supernova gamma-ray burst}
\label{sec:alp_gamma_conv}

\subsection{Core-collapse supernova ALP production}
\label{sec:SNmodels}
  
ALPs are expected to be abundantly produced in CC SNe. 
In a minimal model we consider only their interaction with photons characterized by the Lagrangian term~\cite{Raffelt:1987im}
\begin{equation}
		{\cal L}_{a\gamma}=-\frac{1}{4} \, g_{a\gamma}
		F_{\mu\nu}\tilde{F}^{\mu\nu}a=g_{a\gamma} \, {\bf E}\cdot{\bf B}\,a~,
    \label{eq:lagrangian}
\end{equation}
with $g_{a\gamma}$ the ALP-photon coupling, $F_{\mu\nu}$ the electromagnetic field strength tensor, $\tilde{F}^{\mu\nu}$ its dual, $a$ the ALP field, and $\bE$, $\bB$ the electric and magnetic field, respectively.
This interaction leads to the ALP production rate per volume in the SN core via Primakoff process~\cite{Payez:2014xsa}

\begin{equation}
    \begin{split}
        		\dfrac{d \dot n_a}{dE}&=
		\frac{g_{a\gamma}^{2}\xi^2\, T^3\,E^2}{8\pi^3\, \left( e^{E/T}-1\right) }  \\
		& \left[ \left( 1+\dfrac{\xi^2 T^2}{E^2}\right)\ln\left(1+\frac{E^2}{\xi^2T^2}\right) -1 \right] \,.
	\label{eq:axprod}
    \end{split}
\end{equation}

Here, $E$ is the photon energy measured by a local observer at the emission radius, $T$ the temperature and $\xi^2={\kappa^2}/{4T^2}$ with $\kappa$ the inverse Debye screening length, describing the finite range of the electric field surrounding charged particles in the plasma. The total ALP production rate per unit energy is obtained integrating Eq.~\eqref{eq:axprod} over the SN volume.

In the limit $m_a \ll T$,  the  ALP spectrum is reproduced with excellent precision by the analytical expression~\cite{Payez:2014xsa,Calore:2021hhn}
\begin{equation}
	\frac{dN_{\rm a}}{dE} = C \left(\frac{g_{a\gamma}}{10^{-12}\,\GeV^{-1}}\right)^{2}
	\left(\frac{E}{E_0}\right)^\beta \exp\left( -\frac{(\beta + 1) E}{E_0}\right) \,,
	\label{eq:time-int-spec}
\end{equation}
where the values of the parameters $C$, $E_0$, and $\beta$ are related to the SN model.
The expression above describes a \emph{quasi-thermal} spectrum, with average energy ${\langle E \rangle = E_0}$ and index $\beta$ (in particular, $\beta=2$ corresponds to a perfectly thermal spectrum of ultrarelativistic particles).

\begin{figure}[t!]
	\vspace{0.0cm}
	\includegraphics[width=\columnwidth]{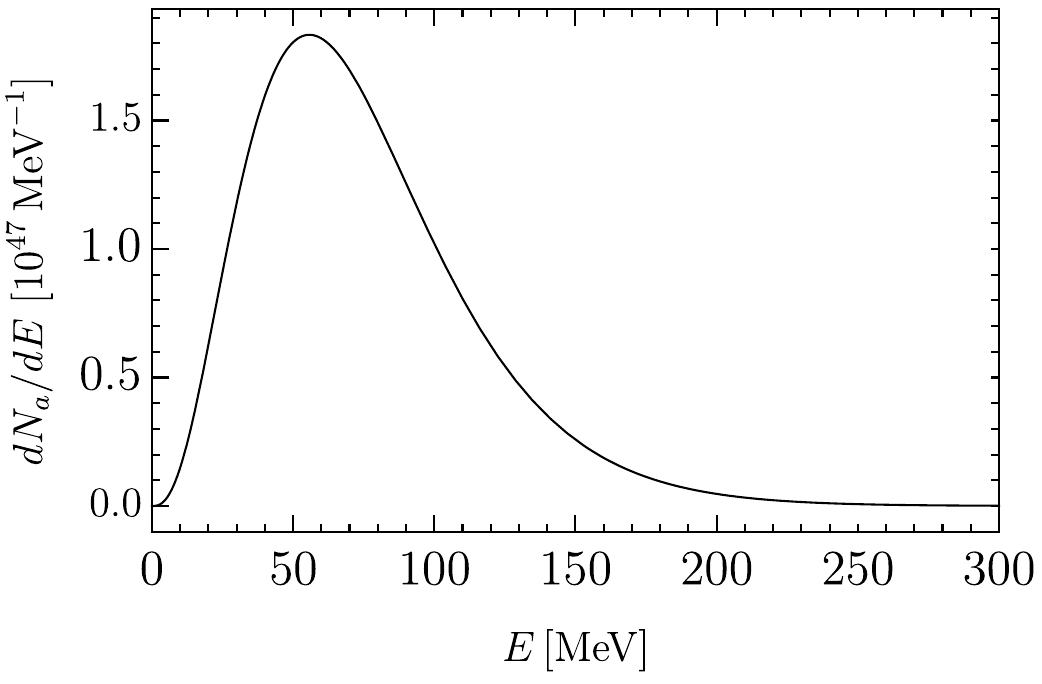}
	\caption{Time-integrated SN ALP spectrum for ${g_{a\gamma}=10^{-12}~\GeV^{-1}}$ and ${C=7.09\times 10^{48}~\MeV^{-1}}$, ${E_0=75.70~\MeV}$ and ${\beta=2.80}$~\cite{Calore:2021hhn}.}
	\label{fig:Phia}
\end{figure}

Following Ref.~\cite{Calore:2021hhn}, it is possible to extract the dependence of the spectral coefficients $C$, $E_0$, and $\beta$ on the SN progenitor mass for successful CC SNe,
\begin{equation}
  \begin{split}
      &\frac{C(M)}{10^{48}\,\MeV^{-1}} =(1.73\pm0.172)\frac{M}{M_{\odot}}-9.74\pm2.92\,,\\
      &\frac{E_0 (M)}{{\rm MeV}} =(1.77\pm0.156)\frac{M}{M_\odot}+59.3\pm2.65\,,\\
      &\beta(M) = (-0.0254\pm 0.00587)\frac{M}{M_\odot} + 2.94 \pm 0.0997\,\,\,.
  \end{split}  
  \label{eq:parameters}
\end{equation}
For our numerical analysis, we will refer to a SN model with an 11.2 $M_{\odot}$ progenitor mass obtained using a 1D spherically symmetric and general relativistic hydrodynamics model, based on the {\tt AGILE BOLTZTRAN} code~\cite{Mezzacappa:1993gn,Liebendoerfer:2002xn}. In this case, an example of the time-integrated spectrum is shown in Fig.~\ref{fig:Phia}, with values of the spectral coefficients taken from Table I in Ref.~\cite{Calore:2021hhn}.

Here, we are interested in studying the ALP detectability prospects, focusing on very weakly interacting ALPs, with $g_{a\gamma}\lesssim10^{-11}\,{\rm GeV^{-1}}$. 
In this regime, ALPs have a very long mean free path even in the very dense SN core, and can escape freely. Yet, even such a tiny ALP-photon coupling may trigger a significant ALP conversion into photons in the Galactic magnetic field, as we will show in the next Section.

In principle, another possible ALP production mechanism in a SN would be the conversion of thermal photons into ALPs in the intense magnetic field inside a SN core, $B\gtrsim\mathcal{O}(10^{9}~{\rm G})$~\cite{Burrows:2007yx,Matsumoto:2020rbz,Mosta:2015ucs,Obergaulinger:2020cqq} (see also Refs.~\cite{Guarini:2020hps,Caputo:2020quz} for studies of ALP-photon conversions in other astrophysical environments, such as the Sun and white dwarfs). However, it can be shown that the light ALP production is unaffected by the possible ALP-photon conversions inside the SN core. Indeed, the photon mean-free-path in this environment is extremely short ($\sim 10^{-8}$~cm) to allow for efficient conversions, that would take place on a length-scale $\sim\mathcal{O}(10^{9}~{\rm cm})$ for $B\sim 10^{9}$~G and $g_{a\gamma}\sim10^{-12}~{\rm GeV}^{-1}$. On the other hand, ALP-photon conversions in a SN core would be relevant for MeV-scale ALPs, which are resonantly produced by the conversion of photons with plasma frequency matching the ALP mass (for transverse photon modes) or energy (for longitudinal photon modes). This process is efficient only in very energetic subclasses of SNe hosting a ultra-high magnetic fields $B\gtrsim10^{14}~{\rm G}$~\cite{Caputo:2021kcv}. Regarding the standard background, the role of strong magnetic fields in the gamma-ray burst (GRB) generation is still under debate but we expect that temporal and spectral information can disentangle it from new physics.

\subsection{Conversion probabilities}
\label{sec:convprob}

The complex structure of the Galactic magnetic field makes the propagation of ALPs in the Milky Way a truly 3-dimensional problem
(see, e.g.,~\cite{DeAngelis:2011id}). 
In this work, we closely follow the technique described in Ref.~\cite{Horns:2012kw} (to which we address the reader for more details) to solve the beam propagation equation along a Galactic line of sight.

To gain some physical intuition, however,  it is also useful to study approximate analytical solutions.
The problem simplifies significantly if the magnetic field ${\bf B}$ is homogeneous. In this case, the  ALP-photon conversion probability is given by~\cite{Raffelt:1987im}
\begin{equation}
P_{a \gamma} 
= (\Delta_{a \gamma} L)^2 \frac{\sin^2(\Delta_{\rm osc} L/2)}{(\Delta_{\rm osc} L/2)^2} \,\ ,
\label{eq:conv}
\end{equation}
where $L$ is the path length, and the oscillation wave number is~\cite{Raffelt:1987im}
\begin{equation}
\Delta_{\rm osc} \equiv \left[(\Delta_{a} - \Delta_{\rm pl})^2 + 4 \Delta_{a \gamma}^2 \right]^{1/2} \,\ ,
\label{eq:deltaosc}
\end{equation}
where  $\Delta_{a}=-m_{a}^{2}/2E$, $\Delta_{\rm pl}=-\omega_{\rm pl}^{2}/2E$ and ${\Delta_{a\gamma}=g_{a\gamma}B_{T}/2}$, being 
$B_T$ the magnetic field component transverse to the photon propagation,
$E$ the  ALP/photon energy, and $\omega_{\rm pl}$ the plasma frequency.
For benchmark values relevant for the SN ALP propagation in the Milky Way the oscillation parameters are
\begin{eqnarray}  
\Delta_{a\gamma}\!\!\! &\simeq &\!\!\! 1.5\times10^{-3} \left(\frac{g_{a\gamma}}{10^{-12}\,\GeV^{-1}} \right)
\left(\frac{B_T}{10^{-6}\,\rm G}\right) \kpc^{-1}
\nonumber\,,\\  
\Delta_a\!\!\! &\simeq &\!\!\!
 -7.8 \times 10^{-5} \left(\frac{m_a}{10^{-11}\, 
        \eV}\right)^2 \left(\frac{E}{100 \,\ \MeV} \right)^{-1} \kpc^{-1}
\nonumber\,,\\  
\Delta_{\rm pl}\!\!\! &\simeq &\!\!\!
 -7.8 \times 10^{-7} \left(\frac{\omega_{\rm pl}}{10^{-12}\, \eV}\right)^2 \left(\frac{E}{100 \,\MeV} \right)^{-1} \kpc^{-1}
\nonumber\, .
\label{eq:Delta0}\end{eqnarray}
Therefore, for the typical conditions that we consider, $\Delta_{\rm pl}\ll\Delta_{a}, \Delta_{a\gamma}$ and thus it is always negligible in our analysis.
The conversion probability in Eq.~\eqref{eq:conv} can be further simplified in the two opposite limits of ${\Delta_a \gg \Delta_{a\gamma}}$ and
$\Delta_a \ll \Delta_{a\gamma}$.
In particular, for $\Delta_a \ll \Delta_{a\gamma}$, corresponding to 
\begin{equation}
\begin{split}
  m_a \ll  m_{a}^{l}=0.044\,{\rm neV}&\left(\frac{g_{a\gamma}}{10^{-12}\,\GeV^{-1}}\right)^{1/2}\\
    &    \left(\frac{B}{10^{-6}\,\textrm{G}}\right)^{1/2}\left(\frac{E}{100 \,\MeV} \right)^{1/2}\,,
    \label{eq:mcrit}
\end{split}    
\end{equation}
 the conversion probability becomes energy independent. 
 Furthermore, for typical Galactic baselines \mbox{$L\lesssim 10~\kpc$} and
 in the conditions of interest for our problem, we can assume $\Delta_{a\gamma} L\ll 1$, which reduces the oscillation probability to
\begin{equation}
P_{a\gamma} \simeq (\Delta_{a \gamma}L)^2 \,\ .
\label{eq:enindep}
\end{equation}
In the opposite limit of $m_{a}\gg m_{a}^{l}$,  $\Delta_{\rm osc} \approx \Delta_a$ and
\begin{equation}
    P_{a\gamma} \simeq 4\left(\frac{\Delta_{a\gamma}}{\Delta_{a}}\right)^{2} {\sin^{2}(\Delta_{a} L/2)}\,,
\end{equation}
which exhibits an energy-dependent oscillatory behavior. 
However, when $\Delta_a\,L \ll 1$, i.e. 
\begin{equation}
   m_a \ll m_{a}^{h}= 0.36\,{\rm neV}\left(\frac{E}{100 \,\ {\rm MeV}} \right)^{1/2}\left(\frac{L}{10 \,\ {\rm kpc}} \right)^{-1/2}\,,
    \label{eq:masslesscond}
\end{equation}
one can expand $\sin^2(\Delta_a\,L/2)\sim \Delta_a^2\,L^2/4$ and the energy-independent probability in Eq.~\eqref{eq:enindep} is recovered. For the conditions considered in this paper, $m_a^h\gg m_a^l$, therefore we dub $m_a^c\equiv m_a^h$ as the critical mass below which the the probability is energy independent. 
Above this threshold, for $m_a\gtrsim m_a^c$, the probability scales as $P_{a\gamma} \sim \Delta_{a}^{-2}$, modulated by fast oscillations that are typically smeared out by the detector finite-energy resolution, see also discussion below.  

\begin{figure}[t!]
	\vspace{0.0cm}
	\includegraphics[width=\columnwidth]{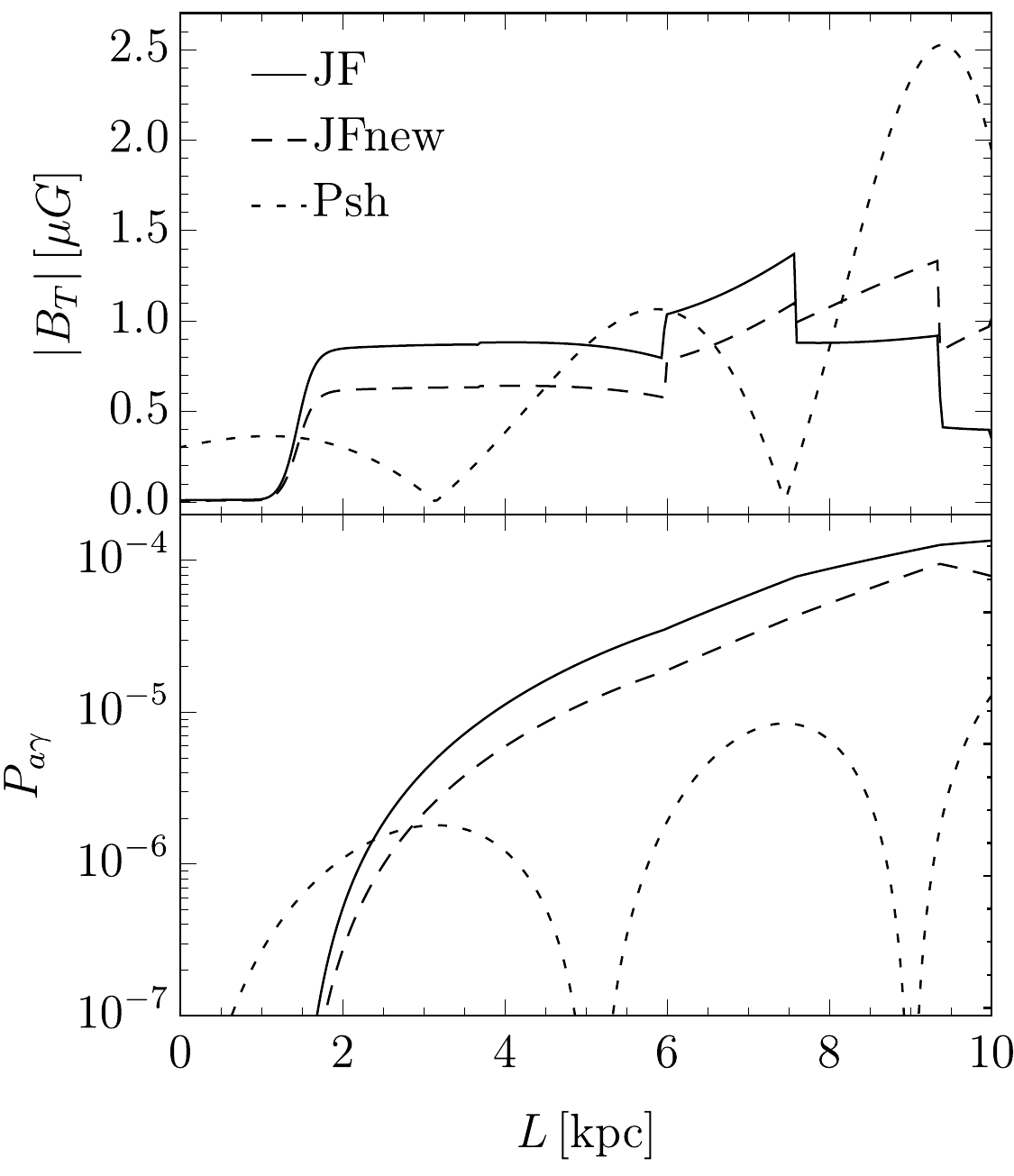}
	\caption{Amplitude of the transverse magnetic field $B_T$ (upper panel) and probability for $m_a\lesssim \mathcal{O}(0.1)$~neV, ${E=100~\MeV}$ and $g_{a\gamma}=10^{-12}~\GeV^{-1}$ (lower panel) as a function of the distance from the source in the direction  ${(\ell, b) = (199.79^{\circ}, -8.96^{\circ})}$ for ``JF'' (black), ``JFnew'' (dashed black) and ``Psh'' (dotted black) models. Note that the source is placed in $L=0$ and the Earth is in $L=10$~kpc.
	}
	\label{fig:Bmodels}
\end{figure}

\begin{figure}[t!]
	\vspace{0.0cm}
	\includegraphics[width=\columnwidth]{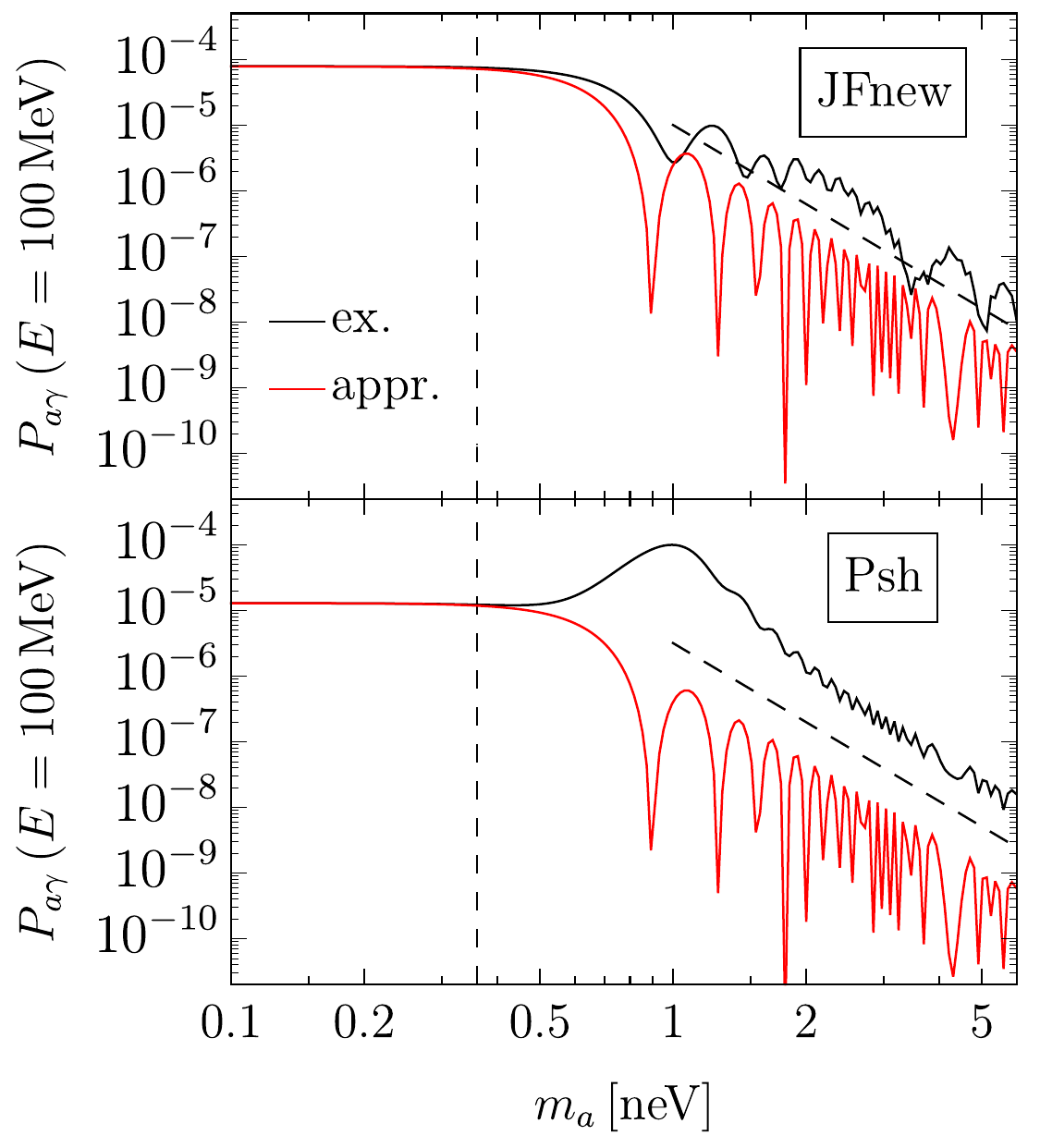}
	\caption{ALP-photon conversion probability $P_{a\gamma}$ for an ALP energy $E=100$~MeV  and ALP-photon coupling ${g_{a\gamma}=10^{-12}~\GeV^{-1}}$ as a function of the ALPs mass $m_{\rm a}$, numerically evaluated for two models of the Galactic magnetic field (black curves) and using the analytical approximation for the average $B$-field (red curves). In the upper panel  the ``JFnew'' model is used, while in the lower panel we refer to the ``Psh'' model. The vertical dashed black line represents transition region around the critical mass $m_a^c \simeq 0.36$~neV at which $\Delta_a\,L = 1$.
The oblique dashed line indicates the behavior $m_{a}^{-4}$, followed by the oscillation probabilities at masses larger than the critical  mass, and it is displayed only to guide the eye (see main text for more details).	
	}
	\label{fig:comparison}
\end{figure}

\begin{figure*}[t!]
	\vspace{0.0cm}
	\includegraphics[scale=0.49]{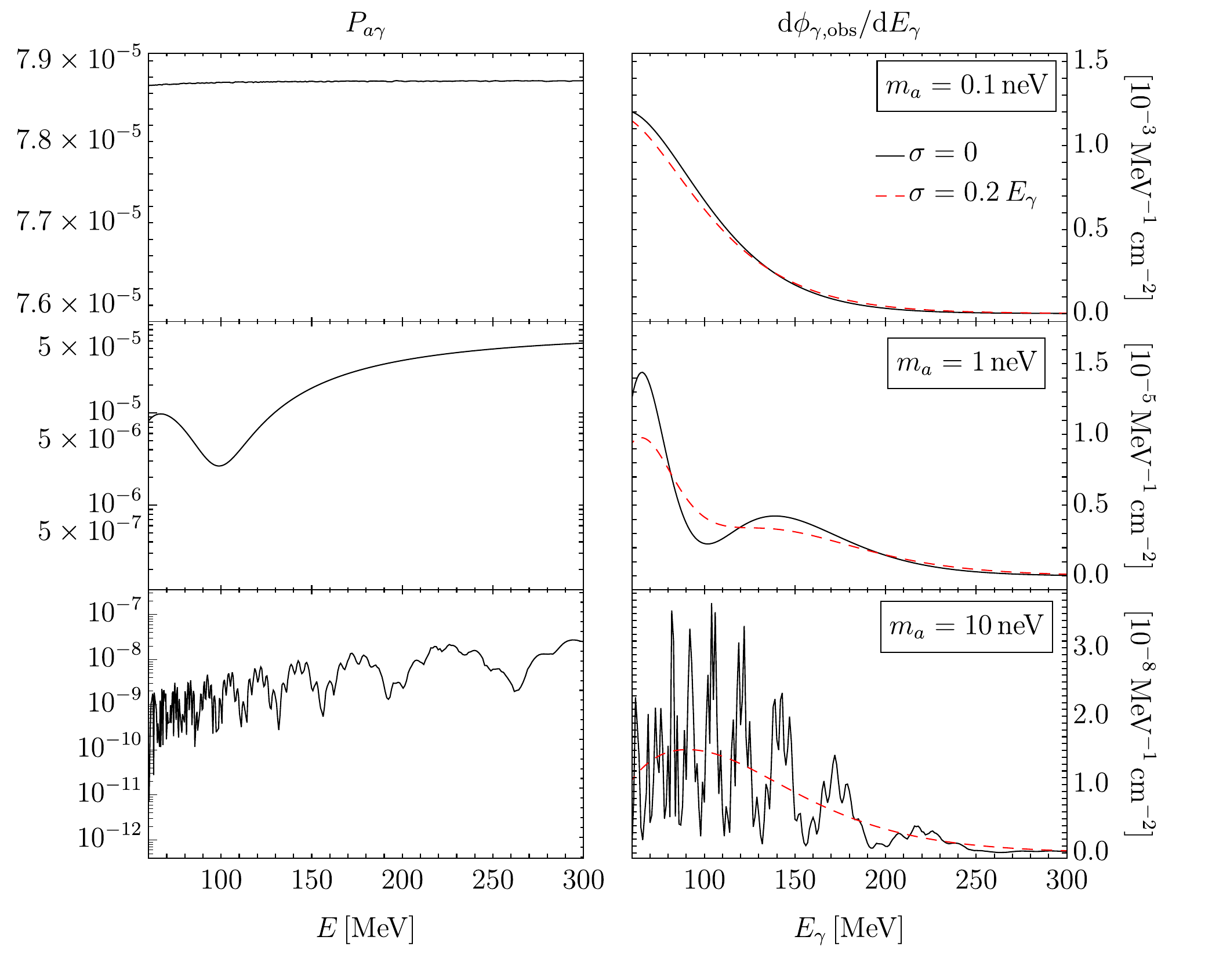}
	\caption{{\it Left panels}: conversion probability $P_{a\gamma}$ as a function of the ALP energy $E$. 
	{\it Right panels}: Observable photon spectra at the detector.  The ALP parameters are $g_{a\gamma}=10^{-12}$~GeV$^{-1}$ and $m_{\rm a}=0.1$~neV (upper panel), $m_{\rm a}=1$~neV (middle panel) or $m_{\rm a}=10$~neV (lower panel). We consider the cases of  perfect energy resolution ($\sigma=0$, black solid lines) and Gaussian energy resolution with $\sigma(E_{\gamma})=0.2\,E_{\gamma}$ (dashed red lines).
	}
	\label{fig:smearing}
\end{figure*}

\begin{figure*}[t!]
\includegraphics[width=\columnwidth]{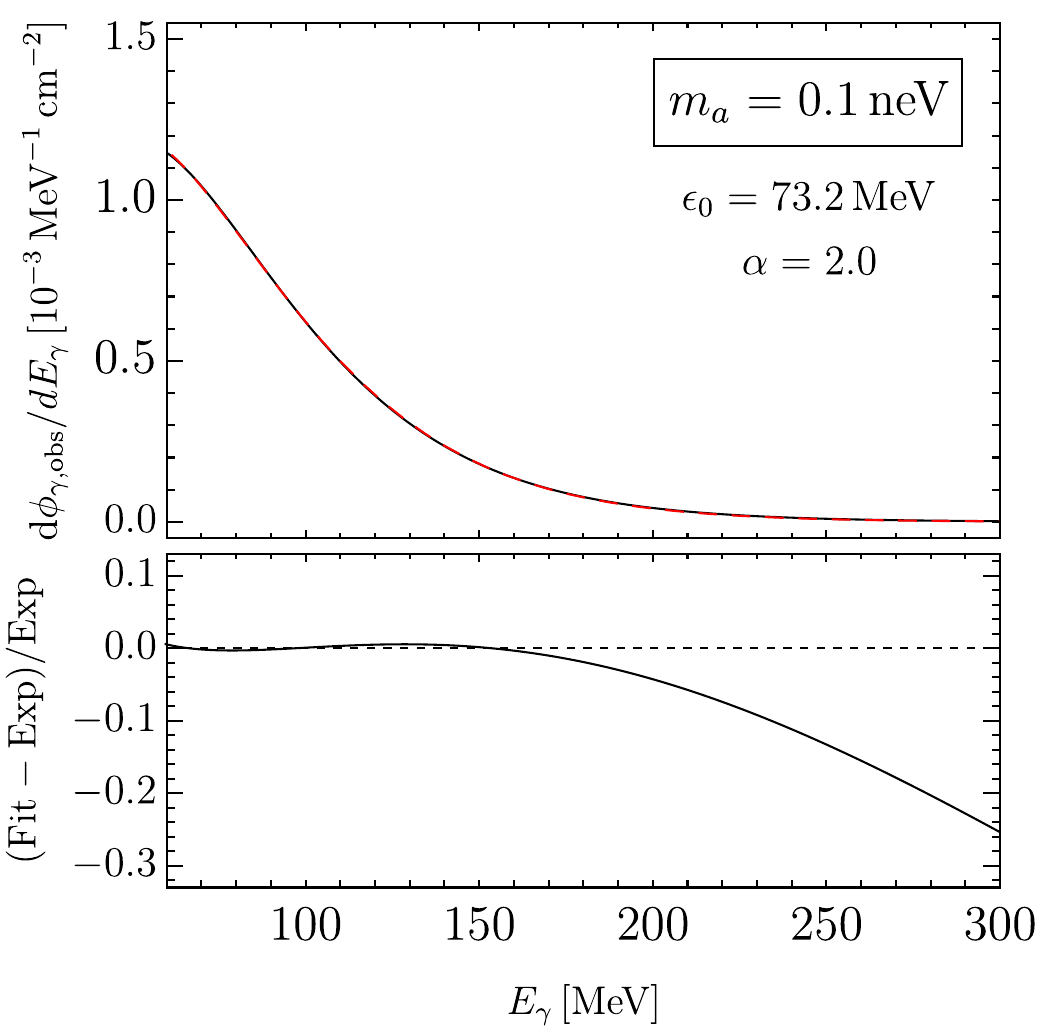}
 \includegraphics[width=\columnwidth]{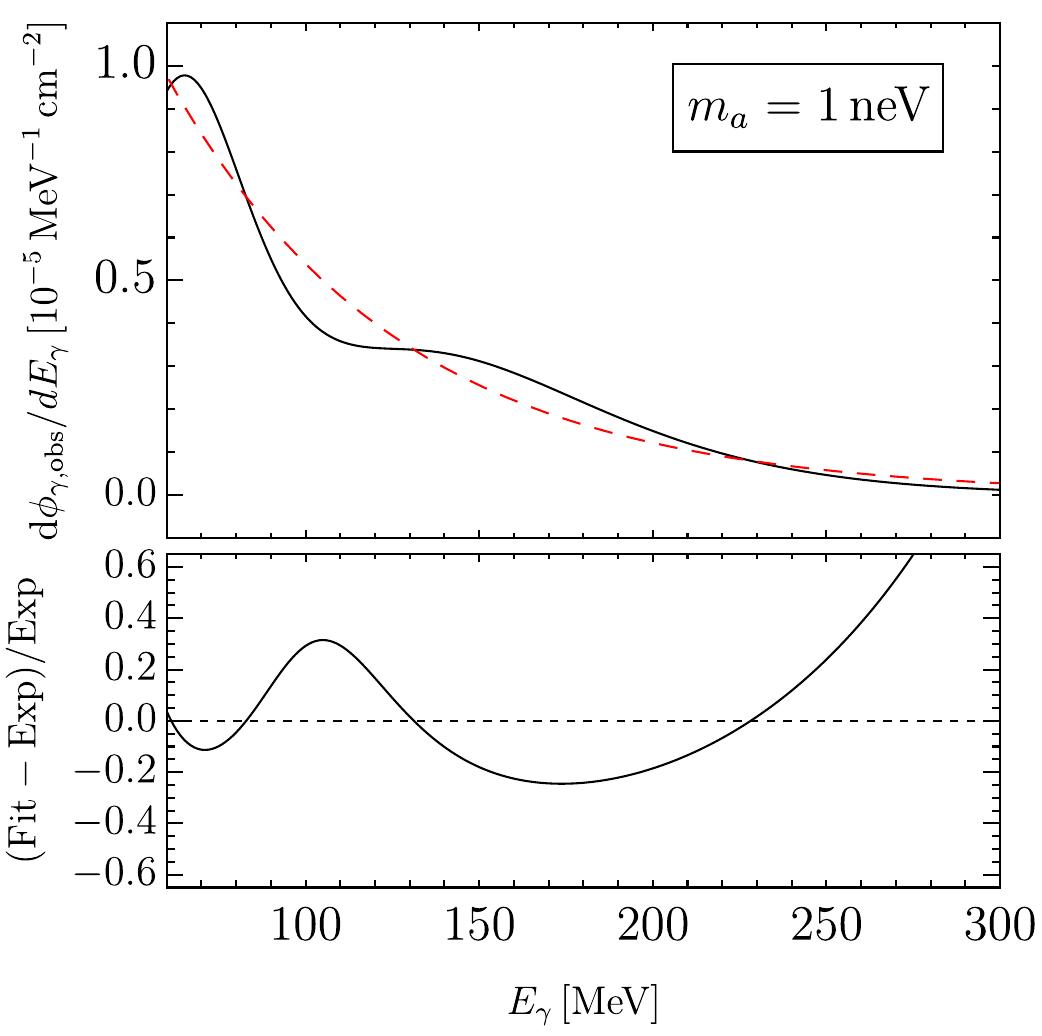}
 
 	\vspace{0.5cm}
 \includegraphics[width=\columnwidth]{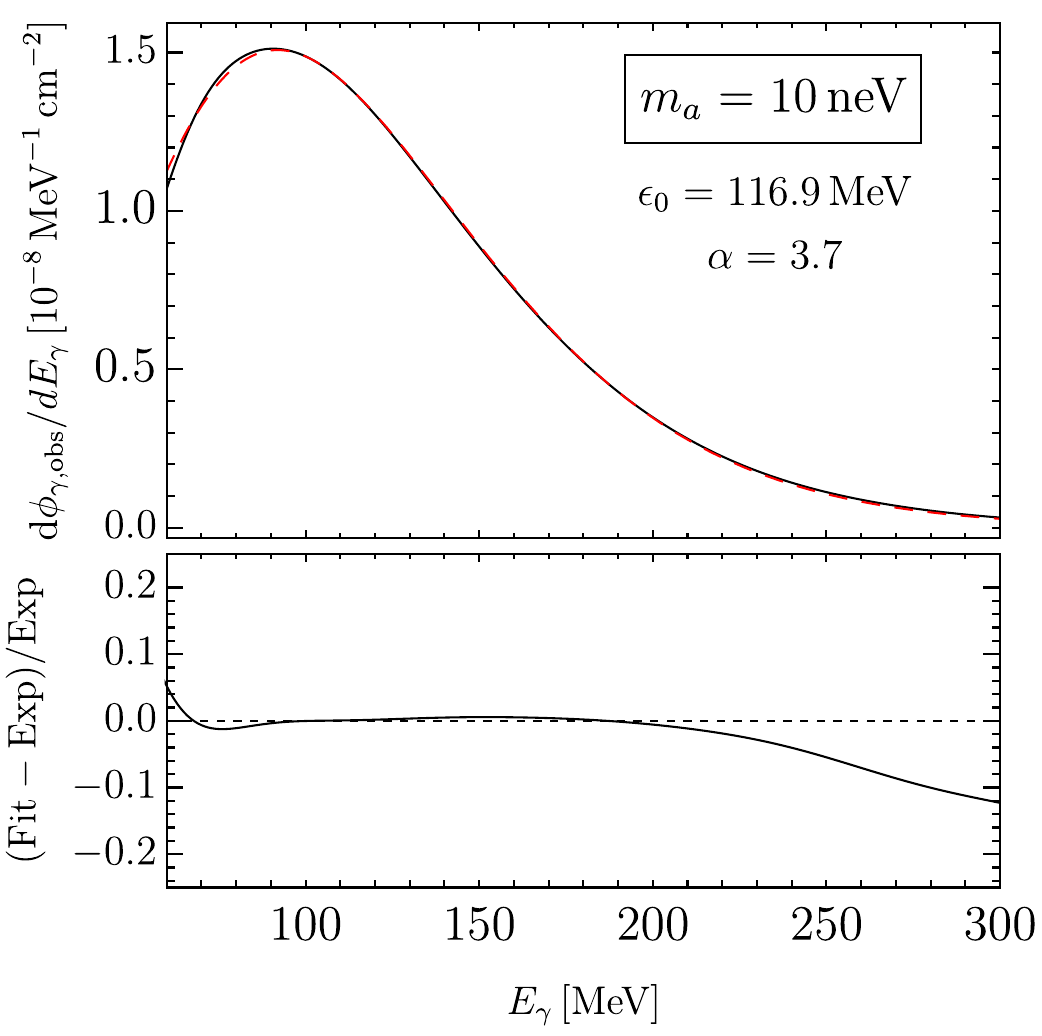}
 
	\caption{Photon spectra expected to be observed (black lines) and their fit (dashed red lines) given by Eq.~\eqref{eq:fitformula} for ${g_{a\gamma}=10^{-12}~\GeV^{-1}}$ and $m_{\rm a}=0.1~\neV$ (upper left), $m_{\rm a}=1~\neV $ (upper right) or $m_{\rm a}=10~\neV$ (lower panel). In each panel, we show also the residuals, defined as the relative difference between the fitted spectrum (Fit) and the expected one (Exp).
 }
	\label{fig:fits}
\end{figure*}

\subsection{Models of Galactic magnetic field}
\label{sec:GMF_models}

Realistic estimates of the ALP-photon conversion probability in the Galaxy rely on the specific morphology of the Milky Way magnetic field. As we shall see, the uncertainties related to the modeling of the Galactic magnetic field contribute significantly to the total uncertainty in our results.
In this work we consider three different state-of-the-art magnetic field models (neglecting their turbulent components~\cite{Carenza:2021alz}), often taken as benchmark in the context of ALP conversions:
\begin{itemize}
    \item ``JF'': Jansson and Farrar model~\cite{Jansson:2012pc}, which includes a disk field and an extended halo field with an out-of-plane component, based on the WMAP7 Galactic Synchrotron Emission map~\cite{Gold:2010fm} and extragalactic Faraday rotation measurements.
    \item ``JFnew'': Jansson and Farrar model~\cite{Jansson:2012pc} with the updated parameters  given in Table~C.2 of~\cite{Adam:2016bgn} (``Jansson12c'' ordered fields), to match the polarized synchrotron and dust emission measured by the Planck satellite~\cite{Planck:2015mrs,Planck:2015qep,Planck:2015zry}. In this work, we use ``JFnew'' as our benchmark model.
    \item ``Psh'': Pshirkov model~\cite{Pshirkov:2011um}, presenting a symmetric (with respect to the Galactic
plane) spiral disk and anti-symmetric halo field, based on rotational measures of extragalactic radio sources~\cite{Kronberg:2009qg}.
\end{itemize}

In the upper panel of Fig.~\ref{fig:Bmodels}, we compare the transverse magnetic field for the different magnetic models listed above in direction $(\ell,b)=(199.79^{\circ},-8.96^{\circ})$,\footnote{\label{foot:betelgeuse}The coordinates used here correspond to direction of Betelgeuse. 
However, this is just to provide a concrete example. Our analysis is intended to be general and not related to a specific star.} over a distance of $10$~kpc from the Earth, i.e.~the source is placed in $L=0$ and the Earth is in $L=10$~kpc. 
For this specific line of sight, 
the ``JF'' (solid line) and ``JFnew'' (dashed line) models predict a transverse magnetic field with a similar shape, where the prediction of ``JFnew'' is larger than ``JF'' for $L>8$~kpc.
On the other hand, the ``Psh''  model (dotted line) leads to a different morphology, featuring a smaller average transverse magnetic field despite the higher values for $L\gtrsim9$~kpc and  $L\lesssim1.5$~kpc compared to the previous models. These different behaviors reflect on the profile of the conversion probability as a function of the distance from the source, as shown in the lower panel of Fig.~\ref{fig:Bmodels} for $m_a\lesssim \mathcal{O}(0.1)$~neV, $E=100$~MeV and $g_{a\gamma}=10^{-12}~\GeV^{-1}$. The value of the probability relevant for observations is the one for $L=10$~kpc,  where the Earth is placed, and it is evident that ``JF'' and ``JFnew'' lead to conversion probabilities of $\mathcal{O}(10^{-4})$, one order of magnitude larger than the one obtained with the ``Psh'' model. 

It is useful to compare the exact conversion probability obtained from a numerical integration of the equations of motion with the analytical expression in Eq.~\eqref{eq:conv}, assuming a uniform magnetic field, obtained averaging the real one over the line of sight (compare with Eq.~(A23) in Ref.~\cite{Mirizzi:2007hr}), i.e.
\begin{equation}
    \langle B_{T}\rangle^2=\frac{1}{L^{2}}\left[{\left|\int_{0}^{L} dz\, B_{x}(z)\right|^{2}+\left|\int_{0}^{L} dz\, B_{y}(z)\right|^{2}}\right]\,,
\label{eq:aveB}
\end{equation}
assuming  ALPs propagating along $z$ over a distance $L$, while 
$B_x$ and $B_y$ are taken from the different magnetic field models. 
This approximation is useful to analyze how the signal is affected by the ALPs mass.
In Fig.~\ref{fig:comparison} we show the conversion probability $P_{a\gamma}$ as a function of the ALP mass $m_{a}$ evaluated by the approximate expression in  Eq.~\eqref{eq:conv} (red line) and compared with the exact result (black line), for $g_{a\gamma}=10^{-12}$~GeV$^{-1}$, $E=100$~MeV  and $L=10$~kpc. 
In the approximate formula we use $\langle B_{T}\rangle=0.58~{\rm \mu G}$ for ``JFnew'' (upper panel), leading to $P_{a\gamma}=7.8\times10^{-5}$ in the low-mass limit, 
$m_a \lesssim 0.36$~neV [see Eq.~\eqref{eq:masslesscond}], indicated  with the vertical dashed line, and $\langle B_{T}\rangle=0.24~{\rm \mu G}$ for ``Psh'' (lower panel), resulting in $P_{a\gamma}=1.3\times10^{-5}$ in the same limit.  
For $m_a \lesssim 0.36$~neV, the approximate and exact results are in agreement since the ALP oscillation length $l_{\rm osc}=2\pi/\Delta_{\rm osc} \approx 3.5 \times 10^3 $~kpc for 
the chosen input values is much larger than the dimension of the Galaxy.  Thus, the conversion probability mainly depends on $\langle B_{T}\rangle$.
For larger masses, the oscillation length becomes smaller (e.g., $l_{\rm osc}\sim 1$~kpc at $m_{a}=3$~neV), making the oscillations  sensitive to the detailed structures of the $B$-field. 
In this latter case, the conversion probabilities evaluated in the exact models are larger than the constant magnetic field approximation. 
This behavior can be explained using the perturbative approach discussed in Ref.~\cite{Marsh:2021ajy}.
In essence, the conversion probability is proportional to the power spectrum of the magnetic field along the line of sight, which is increased by the presence of inhomogeneities. 
Hence, neglecting them in the constant field approximation reduces the oscillation probability.
Nevertheless, in agreement with our previous discussion,
both results exhibit a $m_{a}^{-4}$ behavior, as shown by the dashed line, added in Fig.~\ref{fig:comparison} to guide the eye.

\subsection{Observable gamma-ray spectrum }
\label{sec:gamma-ray-spectrum-analytic}

The gamma-ray flux in units of $\MeV^{-1}\cm^{-2}$ that reaches the Earth is given by

\begin{equation}
    \frac{d\phi_\gamma}{dE}=\frac{1}{4\pi L^2} \frac{dN_{\rm a}}{dE}P_{\rm a \gamma}(E)\;.
    \label{eq:spec1}
\end{equation}

The shape of the observed photon spectrum is determined by 
the finite detector energy resolution,
producing the following smearing

\begin{equation}
    \frac{d\phi_{\gamma,\rm obs}}{dE_\gamma} = \int_{-\infty}^{+\infty} \eta(E,E_\gamma)\frac{d\phi_\gamma}{dE}(E)dE\;,
    \label{eq:spec2}
\end{equation}
where $E$ is the true photon energy,  $E_\gamma$ is the detected one, 
and $\eta(E,E_\gamma)$ is the detector dispersion matrix. 

To illustrate this point, in Fig.~\ref{fig:smearing} we show the observable gamma-ray signal induced by SN ALPs (right panels). The ALP parameters are ${g_{a\gamma}=10^{-12}~\GeV^{-1}}$ and ${m_{ a}=0.1~\neV}$ (upper panel), $m_{a}=1$~neV (middle panel) or  $m_{a}=10$~neV (lower panel). 
We consider the cases of perfect energy resolution ($\sigma=0$, black solid line) and Gaussian energy resolution with $\sigma(E_{\gamma})=0.2\,E_{\gamma}$ (dashed red line). 
The latter is a good approximation for the LAT energy resolution for energies larger than $60~\MeV$, thus we assume a threshold for observation $E_{\rm th}=60~\MeV$.
The SN is assumed to be at a distance of $L=10$~kpc, in the same direction, $(\ell,b)=(199.79^{\circ},-8.96^{\circ})$, specified above, and the magnetic field model is  ``JFnew''.
In the left panels of Fig.~\ref{fig:smearing}, we display the relevant conversions probabilities 
$P_{a \gamma}$ as a function of the ALP energy $E$.

As shown in the upper left panel of Fig.~\ref{fig:smearing}, for \mbox{$m_a=0.1~\neV$} the conversion probability is constant, and thus the effect of the finite-energy resolution does not affect the observed gamma-ray signal, which keeps the same shape of the original ALP spectrum (see Fig.~\ref{fig:Phia}).
Instead, for the other two cases, $P_{a\gamma}$  is energy dependent. 
This would imprint peculiar wiggles in the photon spectrum, which are,  however,  washed out by the finite-energy resolution of the detector (dashed red line in right panels), especially in the $m_{a}=10$~neV case (lower panel). 
Depending on the resolution function, there  may be an intermediate case, e.g. for $m_{a}=1~\neV$ (middle panel), in which also after the energy smearing caused by the finite resolution there is some peculiar energy-dependent modulation of the spectrum. 
However, this is expected to happen only in a very narrow range of ALP masses. 
The discussion above suggests that
there are three different ALP mass ranges with peculiar properties of the expected signal. In absence of wiggles, the observed photon flux in $\MeV^{-1}\,\cm^{-2}$ integrated over the SN explosion time can be very well described by the function
\begin{equation}
\frac{d\phi_{\gamma,\rm obs}}{dE_{\gamma}}= C_{\rm obs}
	\left(\frac{E_{\gamma}}{\epsilon_0}\right)^\alpha \exp\left( -\frac{(\alpha + 1) E_{\gamma}}{\epsilon_0}\right)\,,
	\label{eq:fitformula}
\end{equation}
which resembles the form of the  ALP spectrum in Eq.~\eqref{eq:time-int-spec}, but with different parameters, depending on the considered ALP mass range.

\begin{enumerate}
    \item $m_{a}\ll m_{a}^{c}$. \emph{Energy-independent conversions}. The gamma-ray spectrum has the same shape of the initial  ALP spectrum as in Eq.~\eqref{eq:time-int-spec}.
Thus, comparing it with the original ALP spectrum in Eq.~\eqref{eq:time-int-spec}, the spectral index is $\alpha\simeq \beta$ and the average energy of the observed photon spectrum is $\epsilon_0=\langle E_\gamma \rangle \simeq E_0$.

\item $m_{a} \gg m_{a}^{c}$. \emph{Averaged conversions} with
${P_{a\gamma} \sim 2 \Delta_{a\gamma}^{2}/\Delta_a^{2}}$.
The wiggles in the gamma-ray spectrum induced by ALP conversion  are so dense that they are completely smoothed out by the effect of the detector. 
In this case, $P_{a\gamma} \propto E_{\gamma}^{2}$. 
Thus, the photon spectrum acquires an additional dependence on the energy.
Explicitly, we find
$\alpha=\beta+2$ and $\epsilon_0=\langle E_\gamma \rangle \simeq \frac{\beta+3}{\beta+1} E_0$.

       \item $ m_{a} \approx m_{a}^{c}$. \emph{Intermediate regime}. This is a narrow range where the signal wiggles are not completely washed out by the detector resolution, so we do not expect a smooth functional form in this case and the fit in Eq.~\eqref{eq:fitformula} does not apply.
\end{enumerate}

In Fig.~\ref{fig:fits} we show the results of fitting the three observed spectra from Fig.~\ref{fig:smearing} with Eq.~\eqref{eq:fitformula}.
 The ${m_{a}=0.1~{\rm neV}}$  (upper left panel) is representative of case 1 in which $P_{a \gamma}$ is energy independent, therefore the best-fit parameters $\alpha= 2.0$ and $\epsilon_0=73.2~\MeV$ are close to the average energy $E_{0}$ and spectral index $\beta$ of the \emph{ primary} ALP flux.
 From the plot of the residual it is apparent that the fit worsens in the high-energy tail of the spectrum, at $E_\gamma \gtrsim 150$~MeV, where the discrepancy between the observed flux and the fitted expression is more than $20\%$.

The case of $m_a=1$~neV (upper right panel) corresponds to case 3. Here, due to the energy threshold of the detector, the peak of the energy spectrum is not visible and we can see only the tails. Thus, the fit in Eq.~\eqref{eq:fitformula} does not work and we can try to fit the observed spectrum with an exponential function $d\phi_{\gamma,{\rm obs}}/dE_{\gamma} \propto \exp(-k_1\,E_\gamma)$, with $k_1$ fitting parameter. In this case, the fitting function presents deviations and a remaining energy-dependent modulation of the order of $20\%$ due to  remnant of the oscillatory behavior of the probability after the smearing due to the resolution.

Finally, the plot with $m_a=10$~neV (lower panel) corresponds to case 2. We see that in the energy range $E_\gamma \in [60;200]$~MeV the deviations of the fitting function with respect to the numerical spectrum are less than $10\%$, smaller than the ultralight case due to the larger average energy of the observed photon spectrum. 

We see that the values of the fitting parameters 
$\alpha=3.7$ and $\epsilon_0=116.9~\MeV$ are significantly larger than the ones expected from the original ALP spectrum of Eq.~\eqref{eq:time-int-spec}. 
This feature can be used as a way to distinguish case 3 from case 1. 
Specifically, a value of $\alpha \sim 2$ indicates $m_a\ll m_a^c$, while a larger value, $\alpha \sim 4$, can be considered as evidence of $m_a\gg m_a^c$.

\section{Reconstruction of the signal}
\label{sec:fermi_analysis}

The ALP burst expected from a future Galactic CC SN originates a gamma-ray transient, which falls in the sensitivity range of \fermi-LAT. The LAT enables the detection and re-construction of primary gamma rays of energies from 20 MeV to more than 300 GeV via pair production in its tracker and calorimeter. Per instant of time, it covers a wide field of view of $\sim$2.4 sr \cite{2009ApJ...697.1071A} making it ideal for the study of transient events. In what follows, we assess the potential of the LAT to reconstruct ALP parameters if it detects a gamma-ray burst coming from a Galactic SN explosion during its remaining lifetime. To this end, we rely on simulated data of \fermi-LAT based on the spectra of signal and background components. The simulations are performed with the Fermi Science Tools\footnote{\url{https://fermi.gsfc.nasa.gov/ssc/data/analysis/software/}} (version 2.0.8). Our sensitivity estimate and simulation assume
that the SN transient event will be in the LAT field of view for the entire burst duration.
We therefore do not account for the probability of the LAT to see the SN, and we refer to 
Ref.~\cite{Meyer:2016wrm} for more details.

\subsection{Gamma-ray burst simulation}
\label{sec:data-selection}

We simulate gamma-ray bursts from a Galactic CC SN given by the flux in Eq.~\eqref{eq:spec1}. 
To this end, we will use the fit of the ALP flux obtained from the numerical SN simulation of an $11.2~M_{\odot}$ stellar progenitor at a distance of $L = 10$ kpc in the direction $(\ell, b) = (199.79^{\circ}, -8.96^{\circ})$ [see footnote~\ref{foot:betelgeuse}].
We assume that the gamma-ray burst lasts for $t_{\mathrm{obs}} = 18$ s.

These burst input data are converted into individual gamma-ray events with the Fermi Science Tools routine \texttt{gtobssim} based on the \texttt{FileSpectrum} class.\footnote{See also  
\href{https://fermi.gsfc.nasa.gov/ssc/data/analysis/scitools/obssim_tutorial.html}{fermi.gsfc.nasa.gov} 
for a detailed description of the routine's functionalities.} 
All simulations rely on the \texttt{P8R3\_TRANSIENT020\_V3} event class (Pass8, release 3) and \texttt{FRONT+BACK} event types. We apply further cuts on the selected sample of photons via the requirement of an event zenith angle $<80^{\circ}$, which reduces the contamination of this sample by Earth limb photons, and the additional quality cuts \texttt{DATA\_QUAL>0 \&\& LAT\_CONFIG==1}. We simulate the events in the energy range from 60 MeV to 600 MeV. We do not extend our analysis below 60 MeV since the \textit{Fermi}-LAT effective area is rapidly decreasing at lower energies. Moreover, we consider events arriving from a cone with a radius of $10^{\circ}$ around the position of the SN explosion. Since we cannot simulate \fermi-LAT data in the future, we resort to a typical time interval of duration $t_{\mathrm{obs}}$ in the past observation history of the instrument. For our purposes, such a typical time is characterized by an exposure in the direction $(\ell, b) = (199.79^{\circ}, -8.96^{\circ})$ close to the median value obtained from considering all available time intervals of length $t_{\mathrm{obs}}$. In addition, the explosion is such that it falls completely within a Good Time Interval (GTI) of the LAT. These criteria are fulfilled for $T_{\mathrm{ON}} = 510,160,000$~MET\footnote{Mission elapsed time (MET) is the number of seconds since the reference time of January 1, 2001, at 0h:0m:0s in the Coordinated Universal Time (UTC) system.}, which we adopt as the onset time of the CC SN.

\subsection{Estimating the sensitivity of the LAT to $g_{a\gamma}$}
\label{sec:derive-lat-sensitivity}

One crucial aspect of the ALP burst analysis is the overall sensitivity of the LAT to detect the ALP-induced gamma-ray burst from a CC SN, i.e.~the minimal coupling the LAT will be sensitive to. 
Here, we detail the approach designed to derive the sensitivity for the hypothetical case of an $11.2\,M_{\odot}$ CC SN in the direction ${(\ell, b) = (199.79^{\circ}, -8.96^{\circ})}$.

A suitable statistical framework can be developed from the study presented in~\cite{Meyer:2016wrm}, where the authors analyzed the sensitivity of \textit{Fermi}-LAT to the ALP-flux from CC SN events in the Milky Way and M31. 
Their sensitivity estimates allowed the projection  of upper limits on the ALP parameter space for events occurring within a GTI of the LAT. The approach comprises the following steps:

\begin{enumerate}
\item Select a single day of LAT data when a CC SN is going to occur.
\item Determine all GTIs of this day in the direction of the CC SN (including additional data quality cuts such as zenith angle, Earth limb, etc.).
\item Simulate a CC SN in one of these GTIs.
\item Calculate the LAT exposure for each of these GTIs.
\item Sum the detected events in each GTI except for the one where the CC SN has happened. These are considered ``OFF" counts $N_{\mathrm{OFF, i}}$ and used to create an estimator for the expected background counts $\hat{b}$ in the ``ON" GTI (where the CC SN has exploded). The value of $\hat{b}$ follows from maximizing the Poisson likelihood function
\begin{equation}
\mathcal{L} = \prod_i\frac{(\varepsilon_ib)^{N_{\mathrm{OFF, i}}}}{(N_{\mathrm{OFF, i}})!}e^{-\varepsilon_ib}\,,
\end{equation}
with respect to the background counts $b$,
where $\varepsilon_i = \mathcal{E}_{\mathrm{OFF}, i}/ \mathcal{E}_{\mathrm{ON}}$ is the ratio of exposures in the respective OFF region and ON region (to rescale the background estimator to the OFF exposure since it is going to be determined for the ON-region). It follows that
\begin{equation}
\hat{b} = \frac{\sum_i N_{\mathrm{OFF, i}}}{\sum_i \frac{\mathcal{E}_{\mathrm{OFF}, i}}{\mathcal{E}_{\mathrm{ON}}}}.
\end{equation}
\item Upper limits follow from the ON Poisson likelihood function including the expected average signal counts $s$ derived from the assumed model:
\begin{equation}
\mathcal{L} = \frac{(s + \hat{b})^{N_{\mathrm{ON}}}}{(N_{\mathrm{ON}})!}e^{-(s + \hat{b})}.
\end{equation}
In their case and since there are no true ON counts, they assume a so-called Asimov data set by setting $N_{\mathrm{ON}} = \hat{b}$ (counts must be integers; hence, they take the next integer greater than $\hat{b}$).
\item Upper limits at a certain confidence level are directly constructed from the containment belts following Ref.~\cite{Feldman:1997qc}, which provided a scheme that also works when background and signal are small.
\end{enumerate}

Instead of the total photon counts of a burst, we use a different observable related to the time series of the burst: the time delay among consecutive photons, $\delta\tau_N$, where the index $N$ denotes the size of the group of photons for which the time delay is measured. 
For example, $\delta\tau_5$ means that we consider the difference in arrival time of one selected photon and the 
fifth photon detected after that.
In a real burst of a few seconds (either long or short gamma-ray burst), this time delay should be quite small among the photons associated with the event, while there is no such correlation among photons produced by non-transient events.
Considering $\delta\tau_N$ or the total number of counts is rather equivalent for the case of Galactic gamma-ray bursts since the onset of the transient can be well localized with the help of multi-messenger signals such as neutrinos~\cite{Halzen:2009sm}. 
The advantage of $\delta\tau_N$ is mainly in the context of searches for extragalactic SN explosions (or other transients) where no independent measure of their onset time exists. 
The information in the time series helps  distinguish between upwards fluctuations of photon counts and a genuine transient event characterized by correlated photon arrival times. 
In this work, we examine $\delta\tau_3$, $\delta\tau_4$ and $\delta\tau_5$.

We have modified the approach to consider the time delay discussed above as the ALP observable for determining the LAT sensitivity. We employ the following rationale:
\begin{enumerate}
\item We use a time interval $\Delta t = 30$ s to   derive the time series. In the case of a Galactic CC SN, we will have the neutrino signal as the trigger, which allows us to precisely determine the onset of the burst. This choice of $\Delta t$ ensures that the burst signal is well-contained in the ON bin.
\item We simulate a CC SN in the selected ON region containing $T_{\mathrm{ON}}$ and consider the remaining time intervals as OFF regions. We use real data to determine the background.
\item In the ON region, we simulate the CC SN event multiple times and obtain time series histograms, saving the number of entries in the first time bins (time delay of less than a second/five seconds for $\delta\tau_3$/$\delta\tau_4$/$\delta\tau_5$). We derive the mean expected numbers and repeat for various coupling constant values.
\item We calculate the same histograms and extract the entries in the first bins for each OFF region. We obtain the LAT exposure in the ON and all OFF regions using the Fermi Science Tools routines \texttt{gtselect}, \texttt{gtmktime}, \texttt{gtbin} (LC -- light curve -- with weekly binning), and \texttt{gtexposure}. We derive the estimator for the background counts for each histogram, respectively.
\item We construct containment belts with the method described in \cite{Feldman:1997qc} and derive upper limits for each individual histogram quantity based on the Asimov approach.
\item The final upper limit is the minimal value among the three computed upper limits.
\end{enumerate}

\[\]

\subsection{Reconstructing ALP parameters from time-integrated burst spectra}
\label{sec:parameter_reco_details}

We employ the \textit{Fermi}-LAT data analysis framework of \cite{Muller:2023vjm} to extract model parameters related to ALPs causing the detection of gamma-ray emission from a future nearby SN. Briefly, we fit the model for the time-integrated gamma-ray flux in Eq.~\eqref{eq:fitformula} to mock observations in order to reconstruct the observed parameters $C_{\rm obs}$,
$\alpha$ and $\epsilon_0$ using the generalized Poisson likelihood function
\begin{widetext}
\begin{equation}
\label{eq:poisson_likelihood}
\mathcal{L\!}\left(\left.\bm{\mu} = \bm{S} + \bm{B} + \bm{\delta B}\right|\bm{n}\right) = \prod_{i=1}^{N_E}  \frac{\left(S_{i} + B_i + \delta B_i\right)^{n_{i}}}{\left(n_{i}\right)!}e^{-\left(S_{i} + B_i + \delta B_i\right)}
  \times\exp\!{\left[-\frac{1}{2}\sum_{j,k=1}^{N_E} \delta B_j\,\left(K^{-1}\right)_{jk}\,\delta B_k\right]}\mathrm{.}
\end{equation}
\end{widetext}
The likelihood function involves the ALP-induced gamma-ray emission $\bm{S}$, expected background $\bm{B}$, fluctuations due to finite energy resolution $\bm{\delta B}$, and observed mock data $\bm{n}$ split into $N_E$ energy bins. Background, signal and mock data are simulated and analyzed at the photon counts level. The covariance matrix $K_{ij}$
parameterizes the correlation between spectral fluctuations $\bm{\delta B}$ (associated with both the signal and background), with values obtained from extracting the LAT energy dispersion with respect to the chosen event class using the \emph{Fermi} Science Tools.

The posterior distributions of the model parameters are computed in a Bayesian approach, profiling over nuisance parameters $\bm{\delta B}$ in a maximum likelihood fit with \texttt{iminuit} \cite{iminuit} and employing the profiled log-likelihood function $-2\ln{\mathcal{L}}_{\mathrm{prof}}$ to derive the (marginal) posterior distributions using \texttt{MultiNest} \cite{Feroz:2008xx} (specifying 1000 live points and an evidence tolerance of 0.2). For a detailed description of the method, please refer to our previous publication~\cite{Muller:2023vjm}.

In what follows, we will explore the ALP parameter reconstruction potential of \textit{Fermi}-LAT for couplings $g_{a\gamma}\gtrsim 10^{-12}$ GeV$^{-1}$, well above the detection threshold for ALP-induced bursts given $m_a \lesssim \mathcal{O} (0.1)$ neV (cf.~Fig.~\ref{fig:lat_sensitivity}). 
Hence, we make the simplifying assumption that $\bm{B}\equiv0$. This approximation is justified by the fact that in the short interval of the gamma-ray burst, i.e.~18 s, no substantial background is found. As stated in Ref.~\cite{Meyer:2016wrm}, the mean number of background counts over the full energy range considered here is around 1 for intervals of 20 seconds.

All simulated data does include the time-dependent features of the ALP flux according to the results of the numerical simulations. We describe here the procedure we follow to translate from the time-dependent to the time-integrated quantities.
The time-dependent ALP spectrum is described by
\begin{equation}
\begin{split}
	\frac{dN_{\rm a}}{dE\,dt}& = C(t) \left(\frac{g_{a\gamma}}{10^{-12}\,\GeV^{-1}}\right)^{2}\times\\
	&\hspace{1.cm}
	\left(\frac{E}{E_0 (t)}\right)^{\beta(t)} \exp\left( -\frac{(\beta(t) + 1) E}{E_0 (t)}\right) \,.
\end{split}
\end{equation}
We find that the dependence of the parameters $C$, $\beta$ and $E_0$, for the considered $11.2\,M_{\odot}$ SN model, can be very well approximated by
\begin{equation}
\label{eq:time_evo_spectrum}
    \begin{split}
        C(t)&=\tilde{C}\left(\frac{t}{t_{0}}\right)^{0.545}e^{-0.184t/t_{0}}\,,\\
         E_{0}(t)&=\tilde{E}_{0}\left(\frac{t}{t_{0}}\right)^{0.085}e^{-0.023t/t_{0}}\,,\\
          \beta(t)&=\tilde{\beta}\left(\frac{t}{t_{0}}\right)^{0.048}e^{-0.005t/t_{0}}\,,\\
    \end{split}
\end{equation}
where $t_0 = 1$~s, $\tilde{C} = 6.56\times10^{47}$~MeV$^{-1}$ s$^{-1}$, ${\tilde{E}_{0} = 75.69~\MeV}$ and $\tilde{\beta} = 2.67$. Numerically integrating this time-dependent burst spectrum from 0 to 18 s (in accordance with the simulated gamma-ray burst duration) we obtain the time-integrated spectrum in Eq.~\eqref{eq:time-int-spec}, leading to the time-integrated photon flux fitted with the formula in Eq.~\eqref{eq:fitformula}, employing the energy range from 60 to $600~\MeV$. In our gamma-ray burst simulations we vary $\tilde{C}$, $\tilde{\beta}$ and $\tilde{E}_0$ which -- via this algorithm -- directly translate to the time-integrated quantities $C_{\mathrm{obs}}$, $\alpha$ and $\epsilon_0$. In line with this procedure, we prepare a regular grid for the three parameters $\tilde{C} \in \left[1: 20\right]\times10^{47}~\MeV^{-1}\s^{-1}$, $\tilde{\beta} \in \left[2.5: 2.9\right]$ and $\tilde{E_0} \in \left[68: 82\right]~\MeV$ entering our model $\bm{S}$ for the ALP burst spectrum in Eq.~\eqref{eq:time-int-spec}. We simulate the generated parameter tuples 30 times (different Poisson realizations of the same observation conditions and ALP burst parameters) with \texttt{gtobssim} and take the average to derive the signal model by interpolating the probed data points. Our input (mock) data $\bm{n}$ from which we infer the underlying ALP parameters is a single Poisson realization of a specific parameter set (see Sec.~\ref{sec:results_reconstruction_0}).

\section{Results and discussion}
\label{sec:results}

\subsection{\textit{Fermi}-LAT sensitivity to a Galactic SN}
\label{sec:results_sensitivity}

\begin{figure*}[t!]
	\vspace{0.0cm}
	\includegraphics[width=\textwidth]{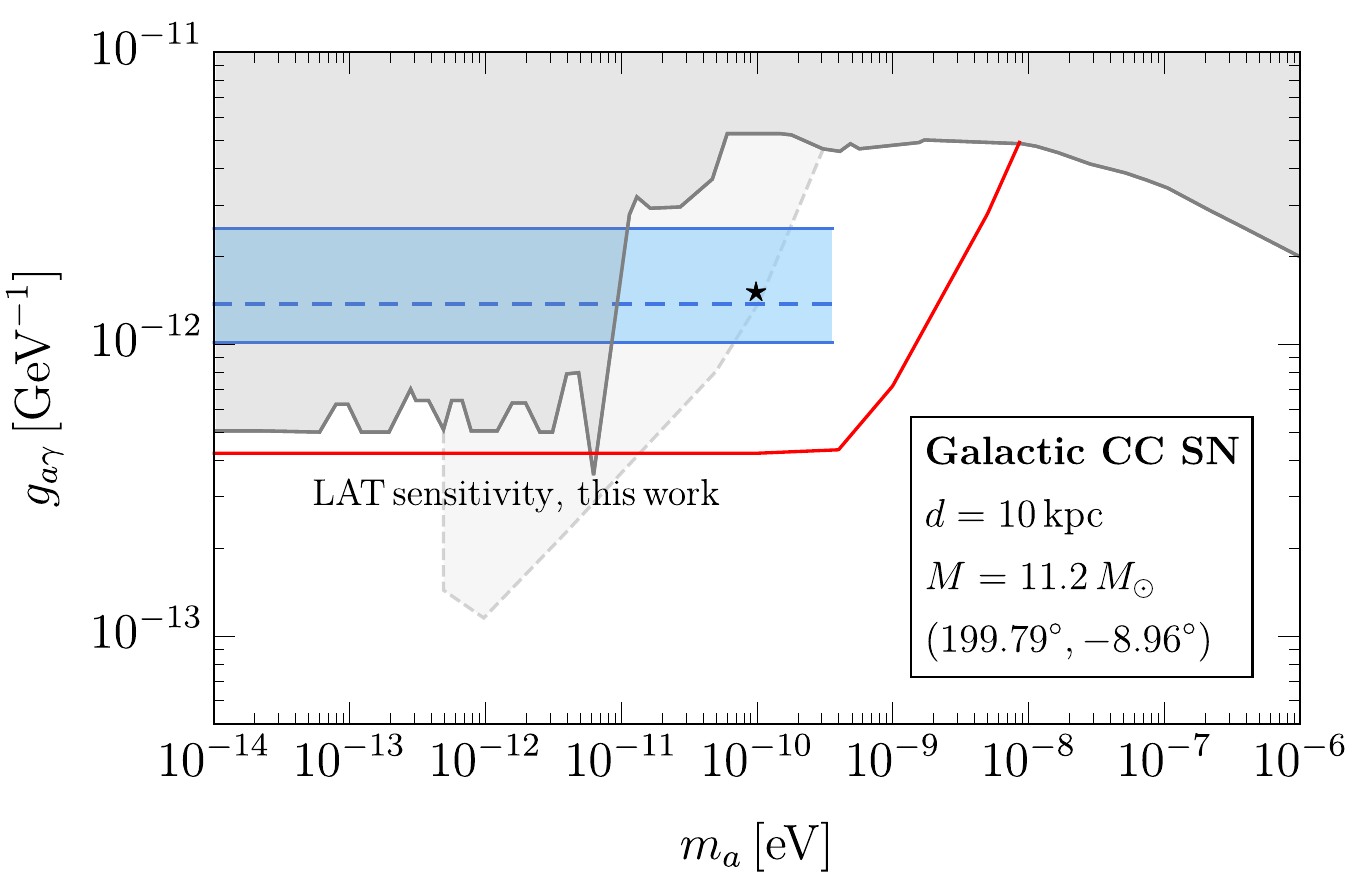}
	\caption{Comparison of the projected 95$\%$ confidence level \textit{Fermi}-LAT sensitivity to $g_{a\gamma}$ (displayed in red)  for an ALP-induced gamma-ray burst from a CC SN with the benchmark characteristics shown in the box and detailed in Sec.~\ref{sec:data-selection}, alongside the current astrophysical constraints	~\cite{Meyer:2016wrm,Reynolds:2019uqt,Dessert:2020lil,Noordhuis:2022ljw,Dessert:2022yqq}, shown in gray (see Ref.~\cite{AxionLimits} and references therein for more details). The black star shows the benchmark ALP parameters $(m_a,g_{a\gamma})=(0.1~\neV,\,1.5\times 10^{-12}~\GeV^{-1})$, discussed in Sec.~\ref{sec:results_reconstruction_0}. The dashed blue line and the light blue band represent the reconstructed coupling and the $1\,\sigma$ error interval, respectively. The light gray region delimited by the dashed line is the axion star explosion limit, obtained assuming ALP dark matter~\cite{Escudero:2023vgv}.
 }
	\label{fig:lat_sensitivity}
\end{figure*}

In Fig.~\ref{fig:lat_sensitivity}, we present the anticipated \textit{Fermi}-LAT sensitivity at 95$\%$ confidence level to $g_{a\gamma}$ for an ALP-induced gamma-ray burst associated with an 11.2 $M_{\odot}$ stellar progenitor SN explosion at a distance $L=10~\kpc$. 
Using the method described in Sec.~\ref{sec:derive-lat-sensitivity}, we find that the LAT could probe currently unexplored ALP masses and couplings. From the LAT data taken in the direction $(\ell, b) = (199.79^{\circ}, -8.96^{\circ})$, we estimate that the average background contribution is $\hat{b}\approx 0$ regarding all three consecutive time-delay observables. Thus, we use $N_{\mathrm{ON}} = 1$ as the background expectation in the selected ``ON'' GTI. From the considered dataset we can also estimate the probability of detecting a SN event in the chosen direction. About one third of the 30 s GTIs exhibit a non-zero exposure of the LAT. It is only in these time frames that a SN explosion can be detected. If we further require that the full burst duration of 18 s is encompassed by the temporal boundaries, than the most optimistic probability to capture such a SN is around $13\%$. However, the LAT can be oriented towards the SN direction in case of a SN alert~\cite{SNEWS:2020tbu}, increasing the probability of a detection.

Despite extensive coverage by other astrophysical probes~\cite{AxionLimits}, the \textit{Fermi}-LAT would explore an unprobed region of the ALP parameter space for $m_a \lesssim 10$ neV, as evident from Fig.~\ref{fig:lat_sensitivity}, where the gray region delimited by the solid line is excluded by astrophysical arguments	~\cite{Meyer:2016wrm,Reynolds:2019uqt,Dessert:2020lil,Noordhuis:2022ljw,Dessert:2022yqq}. 
Hence, a future SN observation by the LAT would significantly improve the ALP parameter space coverage, as already discussed in Ref.~\cite{Meyer:2016wrm}. 
Let us comment, however, that our sensitivity (reaching $g_{a\gamma}\simeq4\times10^{-13}~\GeV^{-1}$) is about a factor of two smaller than the reported sensitivity in Ref.~\cite{Meyer:2016wrm}, which considered a SN at the Galactic center, and hence closer to Earth and in a region where the magnetic field is stronger.
We note that this estimate is influenced by uncertainties in the Galactic magnetic field modeling, which would affect the sensitivity of a factor $\sim 2$ at most, in excess or defect. 
In Fig.~\ref{fig:lat_sensitivity}, the light gray region delimited by the dashed line is constrained by the Planck legacy measurement of the Cosmic Microwave Background optical depth, which can be altered by the explosion of axion stars~\cite{Escudero:2023vgv}. 
Indeed, axion star decays would inject photons into the intergalactic medium (IGM), heating it and leading to an efficient ionization of the IGM, strongly constrained by Planck data~\cite{Planck:2016mks,Planck:2018vyg}. This constraint would supersede the \emph{Fermi}-LAT sensitivity for ALP masses $5\times 10^{-13}~\eV \lesssim m_a \lesssim 10^{-11}~\eV$. However, it is valid only if ALPs are dark matter, an assumption not required in this work. In addition, as further discussed in Ref.~\cite{Escudero:2023vgv}, this bound is largely affected by uncertainties. For instance, it could be modified by different assumptions on core-halo mass relations, still under debate~\cite{Escudero:2023vgv,Mocz:2017wlg,Nori:2020jzx,Mina:2020eik}, and possible ALP self-interactions, which are neglected to obtain the constraint in Fig.~\ref{fig:lat_sensitivity}. For the aforementioned reasons, we neglect this constraint in the following discussion.

\subsection{ALP parameter reconstruction for $m_{\rm a}\lesssim m_{a}^{c}$}
\label{sec:results_reconstruction_0}

Here, we discuss how to reconstruct the ALP-photon coupling in the case of a detection of an ALP-induced gamma-ray burst in coincidence with a future Galactic SN explosion.
We build upon the established \textit{Fermi}-LAT sensitivity to constrain the spectral parameters of the gamma-ray burst when its luminosity is above the derived sensitivity threshold for ALP masses $m_a\lesssim O(0.1)~\neV$, below the critical mass $m_a^c=0.36~\neV$ introduced in Sec.~\ref{sec:gamma-ray-spectrum-analytic}. 
Then, we analyze the integrated spectrum of the burst according to the statistical approach outlined in Sec.~\ref{sec:parameter_reco_details}.

\begin{figure*}[t!]
	\vspace{0.0cm}
	\includegraphics[width=\textwidth]{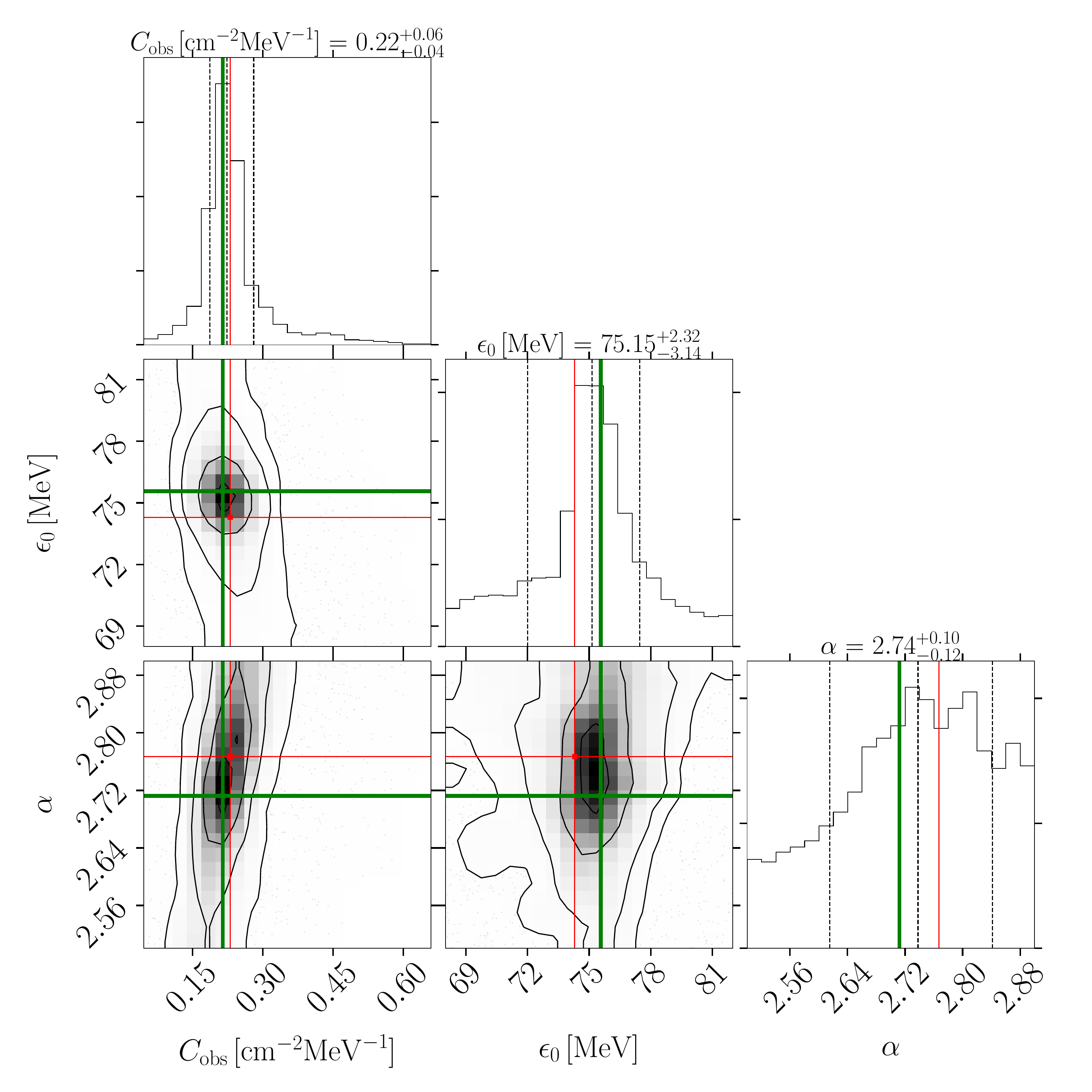}
	\caption{Best-fitting values and posterior distributions of the reconstructed time-integrated ALP spectral parameters $C_{\mathrm{obs}}, \epsilon_0$ and $\alpha$ for an ALP-induced gamma-ray burst of a future SN of an 11.2 $M_{\odot}$ stellar progenitor characterized by $m_a = 0.1$ neV, $g_{a\gamma} = 1.5\times10^{-12}$ GeV$^{-1}$. We overlay the marginal two-dimensional posterior distributions with the parameter values used to simulate the mock signal in red, while the green values denote the parameter values maximizing Eq.~\eqref{eq:poisson_likelihood}. The marginal one-dimensional posterior distributions for each parameter show the $16\%$, $50\%$ (median) and $84\%$ quantiles as dashed black lines, whose numerical values are also stated in the title of each marginal posterior.
 }
	\label{fig:fit_massless_ALP}
\end{figure*}

Specifically, we consider ${m_a = 0.1~\neV}$ and five different coupling strengths ${g_{a\gamma}\in\left[1,\, 1.5,\,2,\, 2.5, \,3\right]\times 10^{-12}~\GeV^{-1}}$ to simulate a mock observation. 
This allows us to comprehensively quantify the uncertainties on the reconstructed spectral parameters. 
In Fig.~\ref{fig:fit_massless_ALP}
we display the case of ${g_{a\gamma} = 1.5\times10^{-12}~\GeV^{-1}}$ as an example of the parameter inference results. 
We find that in all considered scenarios the parameters $\epsilon_0$ and $\alpha$ are reconstructed with a relative uncertainty not higher than $\sim 10\%$. From the results of the statistical analysis shown in Fig.~\ref{fig:fit_massless_ALP}, 
for sake of simplicity, we symmetrize the errors on the flux normalization obtaining the results presented in  Tab.~\ref{tab:results}. 
Although the marginal posterior distribution of $C_{\mathrm{obs}}$ is asymmetric, we adopt the standard deviation of the obtained posterior samples as the proxy for the error on $C_{\mathrm{obs}}$. This procedure underestimates the uncertainty on $g_{a\gamma}$ induced by the statistical inference of the spectral parameters of the burst. 
Yet, as we shall see, the uncertainty on the Galactic magnetic field induces a much larger variance so that our approximation does not strongly bias the final results in Tab.~\ref{tab:results}. 

The observed time-integrated photon flux is fitted by Eq.~\eqref{eq:fitformula} where the normalization $C_{\rm obs}$  can be conveniently expressed as 
\begin{equation}
C_{\rm obs}=  \left(\frac{g_{a\gamma}}{g_{a\gamma,0}}\right)^{4} \left(\frac{B}{B_{0}}\right)^{2} \frac{C_0}{4\pi L^{2}}  P_{a\gamma,0}\,.
\label{eq:Cobs}
\end{equation}
Here, the reference coupling is $g_{a\gamma,0}=10^{-12}~\GeV^{-1}$, ${C_0 \equiv C(M=11.2\,M_{\odot})}$ from Eq.~\eqref{eq:parameters}, and the conversion probability $P_{a\gamma,0}=P_{a\gamma}(g_{a\gamma,0},B_{0})=7.813\times10^{-5}$ is calculated for this given reference coupling and a benchmark magnetic field model characterized by an average transverse magnetic field $B_{0}=0.58~{\rm \mu G}$ along the line of sight. This value is obtained by considering the ``JFnew'' model discussed in Sec.~\ref{sec:alp_gamma_conv}. 
In Tab.~\ref{tab:results}, we report the inferred values of the burst normalization $C_{\rm obs}$ for all five cases.
As expected, the relative uncertainty depends on the total luminosity of the observed burst. Here we note that the central value of the reconstructed $g_{a\gamma}$ is systematically smaller than the real one. This is due to the fact that the fit in Eq.~\eqref{eq:parameters} overestimates the ALP production in the SN for the considered $11.2\,M_{\odot}$ model (see also Fig.~3 in \cite{Calore:2021hhn}).

\begin{table*}[t!]
    \centering
      \caption{Reconstructed (Rec.) ALP-photon coupling with associated error bars (third column, see text for detailed derivation) for different values of the input ALP-photon coupling (first column), leading to different values of $C_{\rm obs}$ (second column).}
    \begin{tabular}{|c|c|c|}
    \hline
          Input $g_{a\gamma}(\times10^{-12}\GeV^{-1})$& $C_{\rm obs}(\MeV^{-1}\cm^{-2})$& Rec. $g_{a\gamma}(\times 10^{-12}\GeV^{-1})$\\
         \hline
        $1.00$& $0.042\pm0.007$& $0.90\substack{+0.76 \\ -0.25}$ \\
        $1.50$& $0.221\pm0.026$&$1.37\substack{+1.12 \\ -0.36}$\\
       $2.00$&  $0.727\pm0.083$&$1.84\substack{+1.50 \\ -0.48}$\\
       $2.50$&  $1.718\pm0.186$&$2.30\substack{+1.87 \\ -0.60}$\\
       $3.00$& $3.758\pm0.384$&$2.78\substack{+2.25 \\ -0.72}$\\
         \hline
    \end{tabular}
    \label{tab:results}
\end{table*}

Inverting Eq.~\eqref{eq:Cobs}, it is possible to reconstruct the ALP-photon coupling for different values of the assumed real value of the coupling. 
The uncertainty on the reconstructed $g_{a\gamma}$ is related to the experimental uncertainty on $C_{\rm obs}$ and the theoretical one on $C_{0}$ as
\begin{equation}
    \frac{\delta g_{a\gamma}}{g_{a\gamma}}=\frac{1}{4}\left(\frac{\delta C_{\rm obs}}{C_{\rm obs}}+\frac{\delta C_0}{C_0}\right)\,\,,
    \label{eq:uncertainty}
\end{equation}
assuming that the distance $L$ is accurately known. 
For the Galactic magnetic field we consider the 3 different models discussed in Sec.~\ref{sec:GMF_models}. The variability related to the different magnetic field models is much larger than the uncertainty evaluated in Eq.~\eqref{eq:uncertainty} fixing the magnetic field.
In Eq.~\eqref{eq:Cobs} $B$ is evaluated as in Eq.~\eqref{eq:aveB} along the line of sight and with $B_{0}=0.58$~${\rm \mu G}$ the benchmark value taken from the ``JFnew'' model.

\begin{table}[t!]
    \centering
        \caption{Reconstructed ALP-photon coupling (third column) for different choices of the Galactic magnetic field model (first column), with average values along the line of sight given in the second column. These values refer to a real coupling $g_{a\gamma}=1.5\times10^{-12}~\GeV^{-1}$.}
    \begin{tabular}{|c|c|c|}
    \hline
         Model & $B~({\rm \mu G})$ & rec. $g_{a\gamma}(\times10^{-12}\GeV^{-1})$\\
         \hline
         ``JF''&$0.76$&$1.19\pm0.19$\\
         ``JFnew''&$0.58$&$1.37\pm0.21$\\
         ``Psh''&$0.235$&$2.15\pm0.33$\\
         \hline
    \end{tabular}
    \label{tab:uncertaintyB}
\end{table}

In Tab.~\ref{tab:uncertaintyB} we show the values of the reconstructed $g_{a\gamma}$ for input value $g_{a\gamma}=1.5\times 10^{-12}~\GeV^{-1}$, assuming different magnetic field models. 
The three models of magnetic field discussed in Sec.~\ref{sec:GMF_models} are listed in decreasing order for the value of the average transverse magnetic field $B$ (second column). 
Clearly, the higher the assumed magnetic field, the lower the reconstructed $g_{a\gamma}$. The Galactic magnetic field is known to be described by the ``JFnew'' model and, if we are confident of this assumption, the uncertainty on $g_{a\gamma}$ goes down to $\pm15\%$. Conservatively, we associate the uncertainty on the reconstructed coupling to the entire range of variability for the three magnetic field models.
The smallest $g_{a\gamma}$ is obtained with the ``JF'' model and the biggest one with the ``Psh'' model. These two extreme values give rise to the uncertainty band on the reconstructed $g_{a\gamma}$.

Following this approach, in Tab.~\ref{tab:results} we show the reconstructed ALP-photon coupling with the associated error bars (third column) for different values of  the ``real'' $g_{a\gamma}$ (first column),  corresponding to a given $C_{\rm obs}$ (second column). The large uncertainty on the magnetic field leads to a $\sim 2$ factor uncertainty on the reconstructed ALP-photon coupling. For clarity, this table can be compared with Tab.~\ref{tab:uncertaintyB}, noticing that the mean value of $g_{a\gamma}$ corresponds to the ``JFnew'' model, which is the most realistic one, and the asymmetric $1\sigma$ error bars are given by the lowest coupling reconstructed with the ``JF'' model and the highest one obtained with the ``Psh'' model. In Fig.~\ref{fig:lat_sensitivity} we show the reconstructed ALP-photon coupling (the dashed blue line) with the $1\,\sigma$ band error associated with the uncertainty on the modeling of the Galactic magnetic field for the benchmark case $m_a=0.1~\neV$ and $g_{a\gamma}=1.5\times 10^{-12}~\GeV^{-1}$. In case of such an observation, the coupling would be reconstructed within a factor two due to the uncertainties on the Galactic magnetic field. On the other hand, a measurement would not reveal any information on the ALP mass. However, for the considered coupling, the strong astrophysical bounds robustly exclude $m_a\lesssim 10^{-11}$~eV. Thus, given the critical mass discussed in Sec.~\ref{sec:gamma-ray-spectrum-analytic}, the observation of an ALP-induced gamma-ray signal in combination with other constraints would allow one to infer an ALP mass $0.01 \neV \lesssim m_a \lesssim \mathcal{O}(0.1)~\neV$.

\subsection{ALP parameter reconstruction for $m_{a} \gtrsim m_{a}^{c}$}
\label{sec:results_reconstruction_massive}

As discussed in Sec.~\ref{sec:alp_gamma_conv}, 
the reconstruction of the ALP-photon coupling in the massive case is less straightforward since the result depends on the unknown ALP mass.

\begin{figure*}[t!]
	\vspace{0.0cm}
 	\includegraphics[width=\columnwidth]{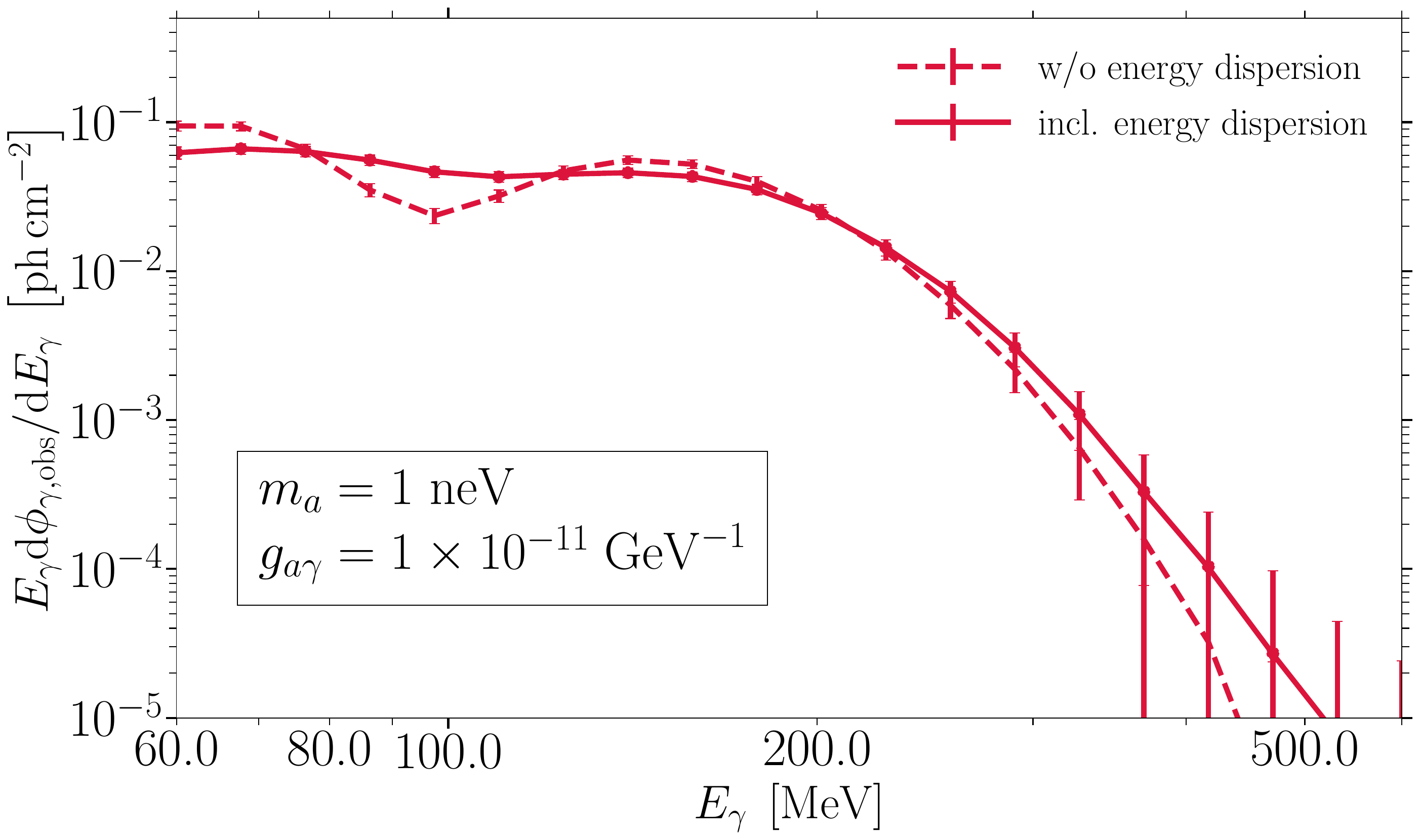}
 		\includegraphics[width=\columnwidth]{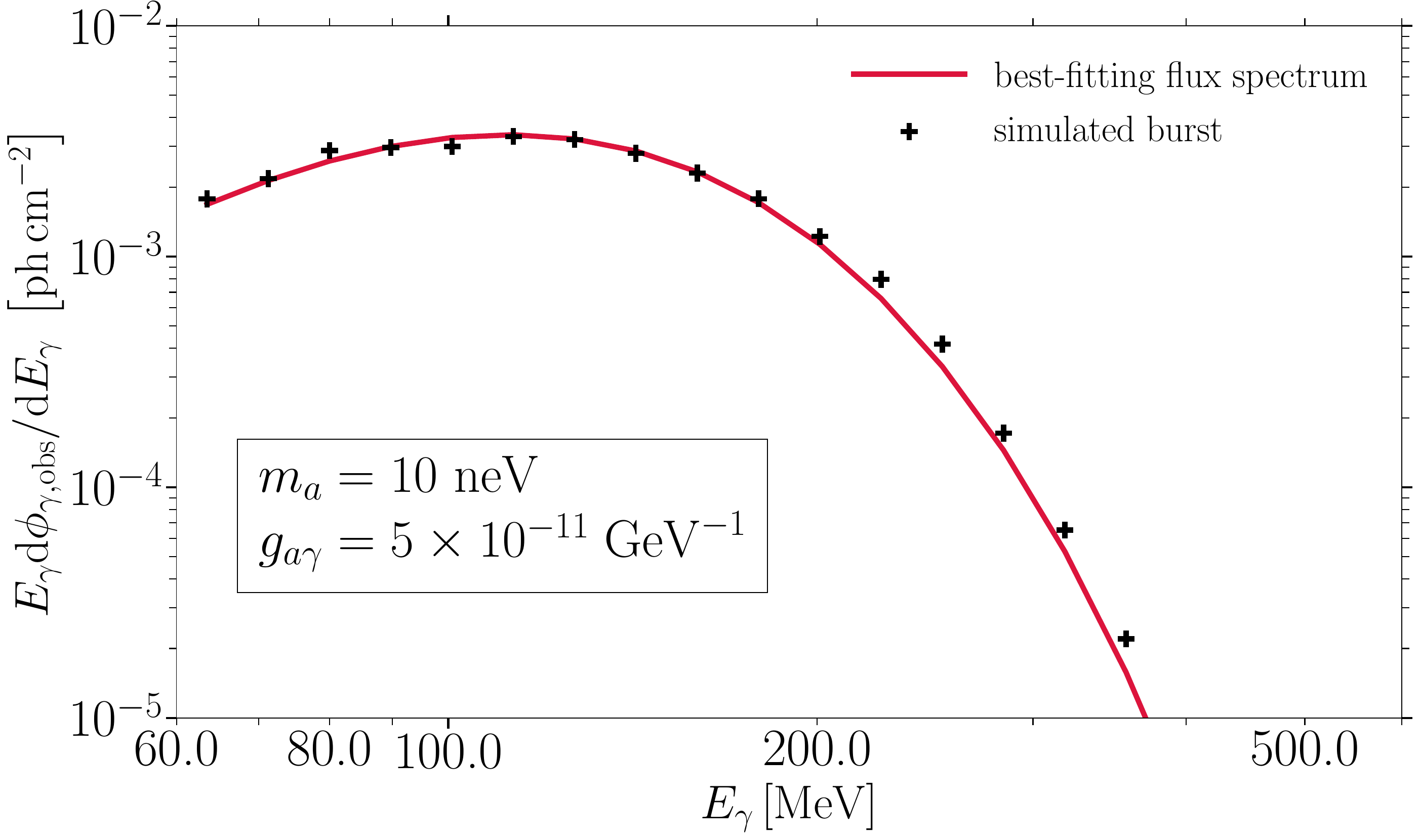}
 	\caption{{\it Left panel}: Simulated gamma-ray spectra between 60 to 600 MeV split in 20 logarithmically spaced energy bins of an ALP burst originating in a CC SN with the benchmark characteristics detailed in Sec.~\ref{sec:data-selection}. The spectra are generated for an ALP mass of $m_a = 1$ neV and $g_{a\gamma} = 10^{-11}$ GeV$^{-1}$. For definiteness, we fix $\beta = 2.77$, $E_0 = 74.3$ MeV as the ALP burst spectrum according to the parameterization in Eq.~\eqref{eq:time-int-spec}. To illustrate the impact of the LAT energy resolution, we distinguish the case with full implementation of the energy dispersion matrix (solid lines) and without any energy dispersion (dashed lines). The error bars indicate the statistical uncertainty due to the Poisson nature of the measurement. The displayed spectra are averaged over 200 Poisson realizations of the signal. \\
 {\it Right panel}: Simulated flux spectrum of a gamma-ray burst induced by ALPs of mass $m_a = 10$ neV and ALP-photon coupling $g_{a\gamma} = 5\times10^{-11}$ GeV$^{-1}$ produced by the same SN stellar progenitor. Black crosses represent the simulated data points, and the red curve represents the fit function based on the analytic description given in Eq.~\eqref{eq:fitformula}. The results have been derived with a $\chi^2$-fit applying the prescription in Eq.~\eqref{eq:chi2_massive}, leading to $\alpha = 4.2$ and $\epsilon_0 = 112~\MeV$.   }
 	
	\label{fig:fit_massive_ALP_LAT}
\end{figure*}

\textbf{Intermediate mass range.} In the intermediate ALP mass range $m_a \approx m_a^c$, the reconstruction of the ALP properties is unfeasible given the performance metrics of the \textit{Fermi}-LAT. Indeed, the ALP-mass induced modulation of the spectrum is almost completely washed out by the poor energy resolution of the detector at these energies. We give a qualitative impression of the impact of the LAT energy dispersion on the modulated time-integrated gamma-ray burst spectrum in the left panel of Fig.~\ref{fig:fit_massive_ALP_LAT}. We illustrate the impact of the LAT energy resolution via a qualitative\footnote{The chosen spectral binning in Fig.~\ref{fig:fit_massive_ALP_LAT} is much finer than the LAT energy resolution would allow for at these energies.} comparison between a simulated ALP burst spectrum (averaged over 200 Poisson realizations) that includes (solid lines) and neglects (dashed lines) the effect of the instrument (irreducible) energy dispersion for the case of $m_a = 1$ neV. 
The impact of the ALP mass is clearly visible for the spectra without energy dispersion, even in this rather coarse binning for a spectral modulation study. This feature is almost completely washed out after including the energy dispersion of \textit{Fermi}-LAT.

\textbf{Large mass range.}
The $m_a \gg m_a^c$ regime can be distinguished from the massless case through the measured value of the $\alpha$ parameter. As discussed in Sec.~\ref{sec:gamma-ray-spectrum-analytic} and illustrated in Fig.~\ref{fig:fits},
the spectral index in this case is much larger than $2$.
We verify that the results of these analytic considerations are retained in a fit to simulated data, i.e.~using the correct LAT instrument response function. To this end, we slightly modify the statistical approach of Sec.~\ref{sec:parameter_reco_details}:
\begin{itemize}
\item We derive the time-integrated flux spectrum of the gamma-ray burst $\bm{n}$ by dividing the simulated photon counts by the exposure per energy bin (utilizing \texttt{gtexposure}). 
\item We obtain the flux $\bm{S}$ from the fit formula in Eq.~\eqref{eq:fitformula} by integration in energy and directly fit its parameters to the observed burst flux.
\item We perform a $\chi^2$-fit with correlated errors based on the covariance matrix $K$ as defined before but now representing flux uncertainties. Namely, we minimize the function:
\begin{equation}
\label{eq:chi2_massive}
    \chi^2 = \left(\bm{n} - \bm{S}\right)^TK^{-1}\left(\bm{n} - \bm{S}\right)\mathrm{.}
\end{equation}
\end{itemize}
In contrast to the case of $m_a = 10$ neV presented in Fig.~\ref{fig:fits}, we select a larger ALP-photon coupling ${g_{a\gamma} = 5\times10^{-11}~\GeV^{-1}}$ to ensure enough statistics. We stress that this coupling strength is already excluded by multiple independent probes so that our results simply showcase what could be done in reality with sufficient statistics. For smaller values of $g_{a\gamma}$, not excluded by other arguments, the statistics is poorer and the reconstruction of the ALP parameters would be more uncertain.

In the right panel of Fig.~\ref{fig:fit_massive_ALP_LAT} we show the results of the $\chi^2$-fit, which yields the best-fitting parameters $\alpha = 4.2$ and ${\epsilon_0 = 112~\MeV}$. The spectral index of $\alpha = 4.2$ confirms the prediction of Sec.~\ref{sec:gamma-ray-spectrum-analytic}, i.e. in the massive case we expect a substantially larger value of the spectral index than for $m_a \ll m_a^c$. 
We emphasize that the derived best-fitting values are subject to large uncertainties of the order of the obtained value itself. Thus, 
though the value of the spectral index can be considered a good indicator of an ALP of mass \emph{above a few neV}, the spectral analysis of the observed gamma-ray burst remains a challenging task.

\section{Conclusions}
\label{sec:conclusions}
In this paper we investigated the possibility of detecting ALPs produced in a nearby SN explosion through their conversion into gamma rays. One may wonder if long GRBs, lasting more than 2 s and partially falling in the same energy range of the ALP-induced signal, could contaminate this signal. Indeed, core-collapse SNe are expected to generate long GRBs associated with jet emission. However, GRB signals are delayed on the order of tens of seconds to
minutes with respect to the core collapse~\cite{2011arXiv1105.4193W,Crnogorcevic:2021wyj}, tagged by the neutrino emission. The origin of such a delay is related to the time needed by the jet to form and propagate through the SN envelope, leading to the GRB production. Therefore, the only source of standard background for the ALP-induced signal would be the gamma-ray emission in the pre-jet phase, whose mechanism is currently poorly understood. In this context, different theoretical models have been proposed to address the origin of such a precursor emission, including photospheric emission model, mini-jets, geometrically thin shells and magnetic reconnection~\cite{Lyutikov:2000qb,MacFadyen:1999mk,Wang:2007nta,Lipunova:2009qf}, featuring temporal and spectral differences. In particular, some models predict a thermal spectrum with average photon energy of $\sim 1~\MeV$ with a time delay of less than 10~s from the core collapse~(see, e.g., Refs.~\cite{Lyutikov:2000qb,Wang:2007nta}). For a Galactic SN this emission may lead to an  isotropic component of high-energy gamma rays, which might fall within the time envelope of the observed neutrino signal. However, given the small amount of observational data, the origin and the properties of such a precursor emission are still under debate~\cite{Lyutikov:2000qb,MacFadyen:1999mk,Wang:2007nta,Lipunova:2009qf}. Therefore, here we have assumed that the ALP-induced signal is not contaminated by any standard physics background and the observed gamma-ray signal would be a hint of new physics.
 
The theoretical framework behind the production and propagation of ALPs has been discussed, and the potential for their detection through the \fermi-LAT has been investigated.

By considering, as a benchmark, a Galactic core-collapse SN at 10 kpc from Earth, we
computed the sensitivity reach of the LAT to be $g_{a\gamma} \simeq 4\times10^{-13}$ GeV$^{-1}$ in the mass range $ m_a \lesssim \mathcal{O}(0.1)~\neV$.
This value can vary by a factor 2 depending on the magnetic field configuration which represents the largest model uncertainty at play, see
discussion in Sec.~\ref{sec:alp_gamma_conv}.

By fully simulating the ALP-induced gamma-ray burst signal with the LAT, we studied how the ALP parameters can be reconstructed 
from the analysis of gamma-ray data. 
We identified three mass ranges which require two different analysis approaches. In particular, in the low-mass regime, 
and high-mass regime, a fit of the gamma-ray signal with the Gamma distribution in Eq.~\eqref{eq:fitformula} can inform us about fundamental properties of the ALPs. 
When the ALPs are lighter than $\mathcal{O}(0.1)$~neV, the reconstruction of the ALP parameters can be very precise, with errors of order 15\%.
Including the variation of the reconstructed parameters because of magnetic field uncertainties, one can reconstruct $g_{a\gamma}$ with errors of about a factor 2. In case of the observation of an ALP-induced photon signal, the interplay with other astrophysical constraints would suggest an ALP mass
$0.01~\neV \lesssim m_a \lesssim \mathcal{O}(0.1)~\neV$.
We also demonstrated that a value of the spectral index $\alpha \sim 4$ will be a good
indicator of an ALP with mass above a few neV.

The results show that a detection of such a signal would provide a unique opportunity to probe the properties of ALPs and shed new light on fundamental physics. 
Here we considered only the case of ALPs coupled with photons but the approach discussed in this work can be easily generalized to ALPs with other interactions, for example to the case of ALPs produced by processes involving nucleons~\cite{Carenza:2019pxu,Carenza:2020cis,Lella:2022uwi,Lella:2023bfb}. 
The detection of gamma-ray signals from SN explosions has been a major topic of interest in astrophysics for several decades, and the potential detection of ALPs adds a new dimension to this field. 
This study highlights the importance of continued research in this area and the potential for exciting new discoveries in the future.

\vspace{0.5cm}

\acknowledgements

We thank Milena Crnogorčević, Manuel Meyer and Edoardo Vitagliano for useful  discussions. This article is based upon work from COST Action COSMIC WISPers CA21106, supported by COST (European Cooperation in Science and Technology). This work has been done thanks to the facilities offered by the Univ. Savoie Mont Blanc - CNRS/IN2P3 MUST computing center. This work is (partially) supported
by ICSC – Centro Nazionale di Ricerca in High Performance Computing,
 Big Data and Quantum Computing, funded by European Union - NextGenerationEU.
The work of P.C. is supported by the European Research Council under Grant No.~742104 and by the Swedish Research Council (VR) under grants  2018-03641 and 2019-02337. 
The work of C.E.~is supported by the ``Agence Nationale de la Recherche'' through grant ANR-19-CE31-0005-01 (PI: F.~Calore). The work of C.~E.~has been supported by the EOSC
Future project which is co-funded by the European Union
Horizon Programme call INFRAEOSC-03-2020, Grant
Agreement 101017536.
The work of A.~M. and G.~L. is partially supported by the Italian Istituto
Nazionale di Fisica Nucleare (INFN) through the ``Theoretical Astroparticle Physics'' project and by the research
grant number 2017W4HA7S ``NAT-NET: Neutrino and
Astroparticle Theory Network'' under the program PRIN
2017 funded by the Italian Ministero dell’Universit\`a e
della Ricerca (MUR).
P.C., M.G. and G.L. thank the Galileo Galilei Institute for Theoretical Physics for 
hospitality during the preparation of part of this work. 

\bibliographystyle{bibi}
\bibliography{biblio.bib}

\providecommand{\noopsort}[1]{}\providecommand{\singleletter}[1]{#1}%

\providecommand{\href}[2]{#2}\begingroup\raggedright\begin{thebibliography}{10}

\bibitem{Grifols:1996id}
J.~A. Grifols, E.~Masso and R.~Toldra, \emph{{Gamma-rays from SN1987A due to
  pseudoscalar conversion}},
  \href{https://doi.org/10.1103/PhysRevLett.77.2372}{\emph{Phys. Rev. Lett.}
  {\bfseries 77} (1996) 2372}
  [\href{https://arxiv.org/abs/astro-ph/9606028}{{\ttfamily
  astro-ph/9606028}}].

\bibitem{Brockway:1996yr}
J.~W. Brockway, E.~D. Carlson and G.~G. Raffelt, \emph{{SN1987A gamma-ray
  limits on the conversion of pseudoscalars}},
  \href{https://doi.org/10.1016/0370-2693(96)00778-2}{\emph{Phys. Lett. B}
  {\bfseries 383} (1996) 439}
  [\href{https://arxiv.org/abs/astro-ph/9605197}{{\ttfamily
  astro-ph/9605197}}].

\bibitem{Payez:2014xsa}
A.~Payez, C.~Evoli, T.~Fischer, M.~Giannotti, A.~Mirizzi and A.~Ringwald,
  \emph{{Revisiting the SN1987A gamma-ray limit on ultralight axion-like
  particles}}, \href{https://doi.org/10.1088/1475-7516/2015/02/006}{\emph{JCAP}
  {\bfseries 02} (2015) 006} [\href{https://arxiv.org/abs/1410.3747}{{\ttfamily
  1410.3747}}].

\bibitem{Jaeckel:2017tud}
J.~Jaeckel, P.~C. Malta and J.~Redondo, \emph{{Decay photons from the axionlike
  particles burst of type II supernovae}},
  \href{https://doi.org/10.1103/PhysRevD.98.055032}{\emph{Phys. Rev. D}
  {\bfseries 98} (2018) 055032}
  [\href{https://arxiv.org/abs/1702.02964}{{\ttfamily 1702.02964}}].

\bibitem{Calore:2020tjw}
F.~Calore, P.~Carenza, M.~Giannotti, J.~Jaeckel and A.~Mirizzi, \emph{{Bounds
  on axionlike particles from the diffuse supernova flux}},
  \href{https://doi.org/10.1103/PhysRevD.102.123005}{\emph{Phys. Rev. D}
  {\bfseries 102} (2020) 123005}
  [\href{https://arxiv.org/abs/2008.11741}{{\ttfamily 2008.11741}}].

\bibitem{Caputo:2021rux}
A.~Caputo, G.~Raffelt and E.~Vitagliano, \emph{{Muonic boson limits: Supernova
  redux}}, \href{https://doi.org/10.1103/PhysRevD.105.035022}{\emph{Phys. Rev.
  D} {\bfseries 105} (2022) 035022}
  [\href{https://arxiv.org/abs/2109.03244}{{\ttfamily 2109.03244}}].

\bibitem{Hoof:2022xbe}
S.~Hoof and L.~Schulz, \emph{{Updated constraints on axion-like particles from
  temporal information in supernova SN1987A gamma-ray data}},
  \href{https://doi.org/10.1088/1475-7516/2023/03/054}{\emph{JCAP} {\bfseries
  03} (2023) 054} [\href{https://arxiv.org/abs/2212.09764}{{\ttfamily
  2212.09764}}].

\bibitem{Muller:2023vjm}
E.~M\"uller, F.~Calore, P.~Carenza, C.~Eckner and M.~C.~D. Marsh,
  \emph{{Investigating the gamma-ray burst from decaying MeV-scale axion-like
  particles produced in supernova explosions}},
  \href{https://doi.org/10.1088/1475-7516/2023/07/056}{\emph{JCAP} {\bfseries
  07} (2023) 056} [\href{https://arxiv.org/abs/2304.01060}{{\ttfamily
  2304.01060}}].

\bibitem{Oberauer:1993yr}
L.~Oberauer, C.~Hagner, G.~Raffelt and E.~Rieger, \emph{{Supernova bounds on
  neutrino radiative decays}},
  \href{https://doi.org/10.1016/0927-6505(93)90004-W}{\emph{Astropart. Phys.}
  {\bfseries 1} (1993) 377}.

\bibitem{Fuller:2008erj}
G.~M. Fuller, A.~Kusenko and K.~Petraki, \emph{{Heavy sterile neutrinos and
  supernova explosions}},
  \href{https://doi.org/10.1016/j.physletb.2008.11.016}{\emph{Phys. Lett. B}
  {\bfseries 670} (2009) 281}
  [\href{https://arxiv.org/abs/0806.4273}{{\ttfamily 0806.4273}}].

\bibitem{Arguelles:2016uwb}
C.~A. Arg\"uelles, V.~Brdar and J.~Kopp, \emph{{Production of keV Sterile
  Neutrinos in Supernovae: New Constraints and Gamma Ray Observables}},
  \href{https://doi.org/10.1103/PhysRevD.99.043012}{\emph{Phys. Rev. D}
  {\bfseries 99} (2019) 043012}
  [\href{https://arxiv.org/abs/1605.00654}{{\ttfamily 1605.00654}}].

\bibitem{DeRocco:2019njg}
W.~DeRocco, P.~W. Graham, D.~Kasen, G.~Marques-Tavares and S.~Rajendran,
  \emph{{Observable signatures of dark photons from supernovae}},
  \href{https://doi.org/10.1007/JHEP02(2019)171}{\emph{JHEP} {\bfseries 02}
  (2019) 171} [\href{https://arxiv.org/abs/1901.08596}{{\ttfamily
  1901.08596}}].

\bibitem{Meyer:2016wrm}
M.~Meyer, M.~Giannotti, A.~Mirizzi, J.~Conrad and M.~A. S\'anchez-Conde,
  \emph{{Fermi Large Area Telescope as a Galactic Supernovae Axionscope}},
  \href{https://doi.org/10.1103/PhysRevLett.118.011103}{\emph{Phys. Rev. Lett.}
  {\bfseries 118} (2017) 011103}
  [\href{https://arxiv.org/abs/1609.02350}{{\ttfamily 1609.02350}}].

\bibitem{Meyer:2020vzy}
M.~Meyer and T.~Petrushevska, \emph{{Search for Axionlike-Particle-Induced
  Prompt $\gamma$-Ray Emission from Extragalactic Core-Collapse Supernovae with
  the $Fermi$ Large Area Telescope}},
  \href{https://doi.org/10.1103/PhysRevLett.124.231101}{\emph{Phys. Rev. Lett.}
  {\bfseries 124} (2020) 231101}
  [\href{https://arxiv.org/abs/2006.06722}{{\ttfamily 2006.06722}}]. [Erratum:
  Phys.Rev.Lett. 125, 119901 (2020)].

\bibitem{Calore:2021hhn}
F.~Calore, P.~Carenza, C.~Eckner, T.~Fischer, M.~Giannotti, J.~Jaeckel,
  K.~Kotake, T.~Kuroda, A.~Mirizzi and F.~Sivo, \emph{{3D template-based
  Fermi-LAT constraints on the diffuse supernova axion-like particle
  background}}, \href{https://doi.org/10.1103/PhysRevD.105.063028}{\emph{Phys.
  Rev. D} {\bfseries 105} (2022) 063028}
  [\href{https://arxiv.org/abs/2110.03679}{{\ttfamily 2110.03679}}].

\bibitem{Raffelt:1987im}
G.~Raffelt and L.~Stodolsky, \emph{{Mixing of the Photon with Low Mass
  Particles}}, \href{https://doi.org/10.1103/PhysRevD.37.1237}{\emph{Phys. Rev.
  D} {\bfseries 37} (1988) 1237}.

\bibitem{Mezzacappa:1993gn}
A.~Mezzacappa and S.~W. Bruenn, \emph{{A numerical method for solving the
  neutrino Boltzmann equation coupled to spherically symmetric stellar core
  collapse}}, \href{https://doi.org/10.1086/172395}{\emph{Astrophys. J.}
  {\bfseries 405} (1993) 669}.

\bibitem{Liebendoerfer:2002xn}
M.~Liebendoerfer, O.~E.~B. Messer, A.~Mezzacappa, S.~W. Bruenn, C.~Y. Cardall
  and F.~K. Thielemann, \emph{{A Finite difference representation of neutrino
  radiation hydrodynamics for spherically symmetric general relativistic
  supernova simulations}},
  \href{https://doi.org/10.1086/380191}{\emph{Astrophys. J. Suppl.} {\bfseries
  150} (2004) 263} [\href{https://arxiv.org/abs/astro-ph/0207036}{{\ttfamily
  astro-ph/0207036}}].

\bibitem{Burrows:2007yx}
A.~Burrows, L.~Dessart, E.~Livne, C.~D. Ott and J.~Murphy, \emph{{Simulations
  of Magnetically-Driven Supernova and Hypernova Explosions in the Context of
  Rapid Rotation}}, \href{https://doi.org/10.1086/519161}{\emph{Astrophys. J.}
  {\bfseries 664} (2007) 416}
  [\href{https://arxiv.org/abs/astro-ph/0702539}{{\ttfamily
  astro-ph/0702539}}].

\bibitem{Matsumoto:2020rbz}
J.~Matsumoto, T.~Takiwaki, K.~Kotake, Y.~Asahina and H.~R. Takahashi, \emph{{2D
  numerical study for magnetic field dependence of neutrino-driven
  core-collapse supernova models}},
  \href{https://doi.org/10.1093/mnras/staa3095}{\emph{Mon. Not. Roy. Astron.
  Soc.} {\bfseries 499} (2020) 4174}
  [\href{https://arxiv.org/abs/2008.08984}{{\ttfamily 2008.08984}}].

\bibitem{Mosta:2015ucs}
P.~M\"osta, C.~D. Ott, D.~Radice, L.~F. Roberts, E.~Schnetter and R.~Haas,
  \emph{{A large scale dynamo and magnetoturbulence in rapidly rotating
  core-collapse supernovae}},
  \href{https://doi.org/10.1038/nature15755}{\emph{Nature} {\bfseries 528}
  (2015) 376} [\href{https://arxiv.org/abs/1512.00838}{{\ttfamily
  1512.00838}}].

\bibitem{Obergaulinger:2020cqq}
M.~Obergaulinger and M.-A. Aloy, \emph{{Magnetorotational core collapse of
  possible GRB progenitors. III. Three-dimensional models}},
  \href{https://doi.org/10.1093/mnras/stab295}{\emph{Mon. Not. Roy. Astron.
  Soc.} {\bfseries 503} (2021) 4942}
  [\href{https://arxiv.org/abs/2008.07205}{{\ttfamily 2008.07205}}].

\bibitem{Guarini:2020hps}
E.~Guarini, P.~Carenza, J.~Galan, M.~Giannotti and A.~Mirizzi,
  \emph{{Production of axionlike particles from photon conversions in
  large-scale solar magnetic fields}},
  \href{https://doi.org/10.1103/PhysRevD.102.123024}{\emph{Phys. Rev. D}
  {\bfseries 102} (2020) 123024}
  [\href{https://arxiv.org/abs/2010.06601}{{\ttfamily 2010.06601}}].

\bibitem{Caputo:2020quz}
A.~Caputo, A.~J. Millar and E.~Vitagliano, \emph{{Revisiting longitudinal
  plasmon-axion conversion in external magnetic fields}},
  \href{https://doi.org/10.1103/PhysRevD.101.123004}{\emph{Phys. Rev. D}
  {\bfseries 101} (2020) 123004}
  [\href{https://arxiv.org/abs/2005.00078}{{\ttfamily 2005.00078}}].

\bibitem{Caputo:2021kcv}
A.~Caputo, P.~Carenza, G.~Lucente, E.~Vitagliano, M.~Giannotti, K.~Kotake,
  T.~Kuroda and A.~Mirizzi, \emph{{Axionlike Particles from Hypernovae}},
  \href{https://doi.org/10.1103/PhysRevLett.127.181102}{\emph{Phys. Rev. Lett.}
  {\bfseries 127} (2021) 181102}
  [\href{https://arxiv.org/abs/2104.05727}{{\ttfamily 2104.05727}}].

\bibitem{DeAngelis:2011id}
A.~De~Angelis, G.~Galanti and M.~Roncadelli, \emph{{Relevance of axion-like
  particles for very-high-energy astrophysics}},
  \href{https://doi.org/10.1103/PhysRevD.84.105030}{\emph{Phys. Rev. D}
  {\bfseries 84} (2011) 105030}
  [\href{https://arxiv.org/abs/1106.1132}{{\ttfamily 1106.1132}}]. [Erratum:
  Phys.Rev.D 87, 109903 (2013)].

\bibitem{Horns:2012kw}
D.~Horns, L.~Maccione, M.~Meyer, A.~Mirizzi, D.~Montanino and M.~Roncadelli,
  \emph{{Hardening of TeV gamma spectrum of AGNs in galaxy clusters by
  conversions of photons into axion-like particles}},
  \href{https://doi.org/10.1103/PhysRevD.86.075024}{\emph{Phys. Rev. D}
  {\bfseries 86} (2012) 075024}
  [\href{https://arxiv.org/abs/1207.0776}{{\ttfamily 1207.0776}}].

\bibitem{Carenza:2021alz}
P.~Carenza, C.~Evoli, M.~Giannotti, A.~Mirizzi and D.~Montanino,
  \emph{{Turbulent axion-photon conversions in the Milky~Way}},
  \href{https://doi.org/10.1103/PhysRevD.104.023003}{\emph{Phys. Rev. D}
  {\bfseries 104} (2021) 023003}
  [\href{https://arxiv.org/abs/2104.13935}{{\ttfamily 2104.13935}}].

\bibitem{Jansson:2012pc}
R.~Jansson and G.~R. Farrar, \emph{{A New Model of the Galactic Magnetic
  Field}}, \href{https://doi.org/10.1088/0004-637X/757/1/14}{\emph{Astrophys.
  J.} {\bfseries 757} (2012) 14}
  [\href{https://arxiv.org/abs/1204.3662}{{\ttfamily 1204.3662}}].

\bibitem{Gold:2010fm}
B.~Gold et~al., \emph{{Seven-Year Wilkinson Microwave Anisotropy Probe (WMAP)
  Observations: Galactic Foreground Emission}},
  \href{https://doi.org/10.1088/0067-0049/192/2/15}{\emph{Astrophys. J. Suppl.}
  {\bfseries 192} (2011) 15} [\href{https://arxiv.org/abs/1001.4555}{{\ttfamily
  1001.4555}}].

\bibitem{Adam:2016bgn}
{\scshape Planck} Collaboration, R.~Adam et~al., \emph{{Planck intermediate
  results.}: {XLII. Large-scale Galactic magnetic fields}},
  \href{https://doi.org/10.1051/0004-6361/201528033}{\emph{Astron. Astrophys.}
  {\bfseries 596} (2016) A103}
  [\href{https://arxiv.org/abs/1601.00546}{{\ttfamily 1601.00546}}].

\bibitem{Planck:2015mrs}
{\scshape Planck} Collaboration, R.~Adam et~al., \emph{{Planck 2015 results. I.
  Overview of products and scientific results}},
  \href{https://doi.org/10.1051/0004-6361/201527101}{\emph{Astron. Astrophys.}
  {\bfseries 594} (2016) A1}
  [\href{https://arxiv.org/abs/1502.01582}{{\ttfamily 1502.01582}}].

\bibitem{Planck:2015qep}
{\scshape Planck} Collaboration, P.~A.~R. Ade et~al., \emph{{Planck 2015
  results - II. Low Frequency Instrument data processings}},
  \href{https://doi.org/10.1051/0004-6361/201525818}{\emph{Astron. Astrophys.}
  {\bfseries 594} (2016) A2}
  [\href{https://arxiv.org/abs/1502.01583}{{\ttfamily 1502.01583}}].

\bibitem{Planck:2015zry}
{\scshape Planck} Collaboration, P.~A.~R. Ade et~al., \emph{{Planck 2015
  results. VI. LFI mapmaking}},
  \href{https://doi.org/10.1051/0004-6361/201525813}{\emph{Astron. Astrophys.}
  {\bfseries 594} (2016) A6}
  [\href{https://arxiv.org/abs/1502.01585}{{\ttfamily 1502.01585}}].

\bibitem{Pshirkov:2011um}
M.~S. Pshirkov, P.~G. Tinyakov, P.~P. Kronberg and K.~J. Newton-McGee,
  \emph{{Deriving global structure of the Galactic Magnetic Field from Faraday
  Rotation Measures of extragalactic sources}},
  \href{https://doi.org/10.1088/0004-637X/738/2/192}{\emph{Astrophys. J.}
  {\bfseries 738} (2011) 192}
  [\href{https://arxiv.org/abs/1103.0814}{{\ttfamily 1103.0814}}].

\bibitem{Kronberg:2009qg}
P.~P. Kronberg and K.~J. Newton-McGee, \emph{{Remarkable symmetries in the
  Milky Way disk's magnetic field}},
  \href{https://doi.org/10.1071/AS10045}{\emph{Publ. Astron. Soc. Austral.}
  {\bfseries 28} (2011) 171} [\href{https://arxiv.org/abs/0909.4753}{{\ttfamily
  0909.4753}}].

\bibitem{Mirizzi:2007hr}
A.~Mirizzi, G.~G. Raffelt and P.~D. Serpico, \emph{{Signatures of Axion-Like
  Particles in the Spectra of TeV Gamma-Ray Sources}},
  \href{https://doi.org/10.1103/PhysRevD.76.023001}{\emph{Phys. Rev. D}
  {\bfseries 76} (2007) 023001}
  [\href{https://arxiv.org/abs/0704.3044}{{\ttfamily 0704.3044}}].

\bibitem{Marsh:2021ajy}
M.~C.~D. Marsh, J.~H. Matthews, C.~Reynolds and P.~Carenza, \emph{{Fourier
  formalism for relativistic axion-photon conversion with astrophysical
  applications}},
  \href{https://doi.org/10.1103/PhysRevD.105.016013}{\emph{Phys. Rev. D}
  {\bfseries 105} (2022) 016013}
  [\href{https://arxiv.org/abs/2107.08040}{{\ttfamily 2107.08040}}].

\bibitem{2009ApJ...697.1071A}
W.~B. {Atwood} et~al., \emph{{The Large Area Telescope on the Fermi Gamma-Ray
  Space Telescope Mission}},
  \href{https://doi.org/10.1088/0004-637X/697/2/1071}{\emph{\apj} {\bfseries
  697} (2009) 1071} [\href{https://arxiv.org/abs/0902.1089}{{\ttfamily
  0902.1089}}].

\bibitem{Feldman:1997qc}
G.~J. Feldman and R.~D. Cousins, \emph{{A Unified approach to the classical
  statistical analysis of small signals}},
  \href{https://doi.org/10.1103/PhysRevD.57.3873}{\emph{Phys. Rev. D}
  {\bfseries 57} (1998) 3873}
  [\href{https://arxiv.org/abs/physics/9711021}{{\ttfamily physics/9711021}}].

\bibitem{Halzen:2009sm}
F.~Halzen and G.~G. Raffelt, \emph{{Reconstructing the supernova bounce time
  with neutrinos in IceCube}},
  \href{https://doi.org/10.1103/PhysRevD.80.087301}{\emph{Phys. Rev. D}
  {\bfseries 80} (2009) 087301}
  [\href{https://arxiv.org/abs/0908.2317}{{\ttfamily 0908.2317}}].

\bibitem{iminuit}
H.~Dembinski, P.~Ongmongkolkul et~al., \emph{scikit-hep/iminuit},  Dec, 2020.
\newblock 10.5281/zenodo.3949207.

\bibitem{Feroz:2008xx}
F.~Feroz, M.~P. Hobson and M.~Bridges, \emph{{MultiNest: an efficient and
  robust Bayesian inference tool for cosmology and particle physics}},
  \href{https://doi.org/10.1111/j.1365-2966.2009.14548.x}{\emph{Mon. Not. Roy.
  Astron. Soc.} {\bfseries 398} (2009) 1601}
  [\href{https://arxiv.org/abs/0809.3437}{{\ttfamily 0809.3437}}].

\bibitem{SNEWS:2020tbu}
{\scshape SNEWS} Collaboration, S.~Al~Kharusi et~al., \emph{{SNEWS 2.0: a
  next-generation supernova early warning system for multi-messenger
  astronomy}}, \href{https://doi.org/10.1088/1367-2630/abde33}{\emph{New J.
  Phys.} {\bfseries 23} (2021) 031201}
  [\href{https://arxiv.org/abs/2011.00035}{{\ttfamily 2011.00035}}].

\bibitem{AxionLimits}
C.~O'Hare, \emph{cajohare/axionlimits: Axionlimits},
  \url{https://cajohare.github.io/AxionLimits/}, July, 2020.
\newblock 10.5281/zenodo.3932430.

\bibitem{Reynolds:2019uqt}
C.~S. Reynolds, M.~C.~D. Marsh, H.~R. Russell, A.~C. Fabian, R.~Smith,
  F.~Tombesi and S.~Veilleux, \emph{{Astrophysical limits on very light
  axion-like particles from Chandra grating spectroscopy of NGC 1275}},
  \href{https://doi.org/10.3847/1538-4357/ab6a0c}{\emph{Astrophys. J.}
  {\bfseries 890} (2020) 59}
  [\href{https://arxiv.org/abs/1907.05475}{{\ttfamily 1907.05475}}].

\bibitem{Dessert:2020lil}
C.~Dessert, J.~W. Foster and B.~R. Safdi, \emph{{X-ray Searches for Axions from
  Super Star Clusters}},
  \href{https://doi.org/10.1103/PhysRevLett.125.261102}{\emph{Phys. Rev. Lett.}
  {\bfseries 125} (2020) 261102}
  [\href{https://arxiv.org/abs/2008.03305}{{\ttfamily 2008.03305}}].

\bibitem{Noordhuis:2022ljw}
D.~Noordhuis, A.~Prabhu, S.~J. Witte, A.~Y. Chen, F.~Cruz and C.~Weniger,
  \emph{{Novel Constraints on Axions Produced in Pulsar Polar-Cap Cascades}},
  \href{https://doi.org/10.1103/PhysRevLett.131.111004}{\emph{Phys. Rev. Lett.}
  {\bfseries 131} (2023) 111004}
  [\href{https://arxiv.org/abs/2209.09917}{{\ttfamily 2209.09917}}].

\bibitem{Dessert:2022yqq}
C.~Dessert, D.~Dunsky and B.~R. Safdi, \emph{{Upper limit on the axion-photon
  coupling from magnetic white dwarf polarization}},
  \href{https://doi.org/10.1103/PhysRevD.105.103034}{\emph{Phys. Rev. D}
  {\bfseries 105} (2022) 103034}
  [\href{https://arxiv.org/abs/2203.04319}{{\ttfamily 2203.04319}}].

\bibitem{Escudero:2023vgv}
M.~Escudero, C.~K. Pooni, M.~Fairbairn, D.~Blas, X.~Du and D.~J.~E. Marsh,
  \emph{{Axion Star Explosions: A New Source for Axion Indirect Detection}},
  \href{https://arxiv.org/abs/2302.10206}{{\ttfamily 2302.10206}}.

\bibitem{Planck:2016mks}
{\scshape Planck} Collaboration, R.~Adam et~al., \emph{{Planck intermediate
  results. XLVII. Planck constraints on reionization history}},
  \href{https://doi.org/10.1051/0004-6361/201628897}{\emph{Astron. Astrophys.}
  {\bfseries 596} (2016) A108}
  [\href{https://arxiv.org/abs/1605.03507}{{\ttfamily 1605.03507}}].

\bibitem{Planck:2018vyg}
{\scshape Planck} Collaboration, N.~Aghanim et~al., \emph{{Planck 2018 results.
  VI. Cosmological parameters}},
  \href{https://doi.org/10.1051/0004-6361/201833910}{\emph{Astron. Astrophys.}
  {\bfseries 641} (2020) A6}
  [\href{https://arxiv.org/abs/1807.06209}{{\ttfamily 1807.06209}}]. [Erratum:
  Astron.Astrophys. 652, C4 (2021)].

\bibitem{Mocz:2017wlg}
P.~Mocz, M.~Vogelsberger, V.~H. Robles, J.~Zavala, M.~Boylan-Kolchin,
  A.~Fialkov and L.~Hernquist, \emph{{Galaxy formation with BECDM \textendash{}
  I. Turbulence and relaxation of idealized haloes}},
  \href{https://doi.org/10.1093/mnras/stx1887}{\emph{Mon. Not. Roy. Astron.
  Soc.} {\bfseries 471} (2017) 4559}
  [\href{https://arxiv.org/abs/1705.05845}{{\ttfamily 1705.05845}}].

\bibitem{Nori:2020jzx}
M.~Nori and M.~Baldi, \emph{{Scaling relations of fuzzy dark matter haloes
  \textendash{} I. Individual systems in their cosmological environment}},
  \href{https://doi.org/10.1093/mnras/staa3772}{\emph{Mon. Not. Roy. Astron.
  Soc.} {\bfseries 501} (2021) 1539}
  [\href{https://arxiv.org/abs/2007.01316}{{\ttfamily 2007.01316}}].

\bibitem{Mina:2020eik}
M.~Mina, D.~F. Mota and H.~A. Winther, \emph{{Solitons in the dark: First
  approach to non-linear structure formation with fuzzy dark matter}},
  \href{https://doi.org/10.1051/0004-6361/202038876}{\emph{Astron. Astrophys.}
  {\bfseries 662} (2022) A29}
  [\href{https://arxiv.org/abs/2007.04119}{{\ttfamily 2007.04119}}].

\bibitem{2011arXiv1105.4193W}
S.~E. {Woosley}, \emph{{Models for Gamma-Ray Burst Progenitors and Central
  Engines}}, \href{https://doi.org/10.48550/arXiv.1105.4193}{\emph{arXiv
  e-prints} (2011) arXiv:1105.4193}
  [\href{https://arxiv.org/abs/1105.4193}{{\ttfamily 1105.4193}}].

\bibitem{Crnogorcevic:2021wyj}
M.~Crnogor\v{c}evi\'c, R.~Caputo, M.~Meyer, N.~Omodei and M.~Gustafsson,
  \emph{{Searching for axionlike particles from core-collapse supernovae with
  Fermi LAT\textquoteright{}s low-energy technique}},
  \href{https://doi.org/10.1103/PhysRevD.104.103001}{\emph{Phys. Rev. D}
  {\bfseries 104} (2021) 103001}
  [\href{https://arxiv.org/abs/2109.05790}{{\ttfamily 2109.05790}}].

\bibitem{Lyutikov:2000qb}
M.~Lyutikov and V.~Usov, \emph{{Precursors of gamma-ray bursts: a clue to the
  burster's nature}}, \href{https://doi.org/10.1086/317278}{\emph{Astrophys. J.
  Lett.} {\bfseries 543} (2000) L129}
  [\href{https://arxiv.org/abs/astro-ph/0007291}{{\ttfamily
  astro-ph/0007291}}].

\bibitem{MacFadyen:1999mk}
A.~I. MacFadyen, S.~E. Woosley and A.~Heger, \emph{{Supernovae, jets, and
  collapsars}}, \href{https://doi.org/10.1086/319698}{\emph{Astrophys. J.}
  {\bfseries 550} (2001) 410}
  [\href{https://arxiv.org/abs/astro-ph/9910034}{{\ttfamily
  astro-ph/9910034}}].

\bibitem{Wang:2007nta}
X.-Y. Wang and P.~Meszaros, \emph{{GRB Precursors in the Fallback Collapsar
  Scenario}}, \href{https://doi.org/10.1086/522820}{\emph{Astrophys. J.}
  {\bfseries 670} (2007) 1247}
  [\href{https://arxiv.org/abs/astro-ph/0702441}{{\ttfamily
  astro-ph/0702441}}].

\bibitem{Lipunova:2009qf}
G.~V. Lipunova, E.~S. Gorbovskoy, A.~I. Bogomazov and V.~M. Lipunov,
  \emph{{Population synthesis of gamma-ray bursts with precursor activity and
  the spinar paradigm}},
  \href{https://doi.org/10.1111/j.1365-2966.2009.15079.x}{\emph{Mon. Not. Roy.
  Astron. Soc.} {\bfseries 397} (2009) 1695}
  [\href{https://arxiv.org/abs/0903.3169}{{\ttfamily 0903.3169}}].

\bibitem{Carenza:2019pxu}
P.~Carenza, T.~Fischer, M.~Giannotti, G.~Guo, G.~Mart\'\i{}nez-Pinedo and
  A.~Mirizzi, \emph{{Improved axion emissivity from a supernova via
  nucleon-nucleon bremsstrahlung}},
  \href{https://doi.org/10.1088/1475-7516/2019/10/016}{\emph{JCAP} {\bfseries
  10} (2019) 016} [\href{https://arxiv.org/abs/1906.11844}{{\ttfamily
  1906.11844}}]. [Erratum: JCAP 05, E01 (2020)].

\bibitem{Carenza:2020cis}
P.~Carenza, B.~Fore, M.~Giannotti, A.~Mirizzi and S.~Reddy, \emph{{Enhanced
  Supernova Axion Emission and its Implications}},
  \href{https://doi.org/10.1103/PhysRevLett.126.071102}{\emph{Phys. Rev. Lett.}
  {\bfseries 126} (2021) 071102}
  [\href{https://arxiv.org/abs/2010.02943}{{\ttfamily 2010.02943}}].

\bibitem{Lella:2022uwi}
A.~Lella, P.~Carenza, G.~Lucente, M.~Giannotti and A.~Mirizzi,
  \emph{{Protoneutron stars as cosmic factories for massive axionlike
  particles}}, \href{https://doi.org/10.1103/PhysRevD.107.103017}{\emph{Phys.
  Rev. D} {\bfseries 107} (2023) 103017}
  [\href{https://arxiv.org/abs/2211.13760}{{\ttfamily 2211.13760}}].

\bibitem{Lella:2023bfb}
A.~Lella, P.~Carenza, G.~Co', G.~Lucente, M.~Giannotti, A.~Mirizzi and
  T.~Rauscher, \emph{{Getting the most on supernova axions}},
  \href{https://doi.org/10.1103/PhysRevD.109.023001}{\emph{Phys. Rev. D}
  {\bfseries 109} (2024) 023001}
  [\href{https://arxiv.org/abs/2306.01048}{{\ttfamily 2306.01048}}].

\end{thebibliography}\endgroup

\end{document}